\RequirePackage{fix-cm}
\ifdefined\pdfminorversion\pdfminorversion=7\fi
\documentclass{article}
\usepackage[T1]{fontenc}
\usepackage{iclr2027_conference,times}

\usepackage{amsmath,amsfonts,bm}

\def\eqref#1{equation~\ref{#1}}

\def\floor#1{\lfloor #1 \rfloor}
\def\1{\bm{1}}

\DeclareMathAlphabet{\mathsfit}{\encodingdefault}{\sfdefault}{m}{sl}
\SetMathAlphabet{\mathsfit}{bold}{\encodingdefault}{\sfdefault}{bx}{n}

\def\sA{{\mathbb{A}}}
\def\sB{{\mathbb{B}}}
\def\sC{{\mathbb{C}}}
\def\sD{{\mathbb{D}}}

\def\sI{{\mathbb{I}}}

\def\sK{{\mathbb{K}}}

\def\sO{{\mathbb{O}}}
\def\sP{{\mathbb{P}}}

\def\sS{{\mathbb{S}}}
\def\sT{{\mathbb{T}}}

\def\sV{{\mathbb{V}}}
\def\sW{{\mathbb{W}}}

\DeclareMathOperator*{\argmax}{arg\,max}

\usepackage[hidelinks]{hyperref}
\usepackage{xurl}
\usepackage{ragged2e}
\usepackage{microtype}
\usepackage{graphicx}
\usepackage{booktabs}
\usepackage{multirow}
\usepackage{tabularx}
\usepackage{longtable}
\usepackage{algorithm}
\usepackage{algpseudocode}
\usepackage{amsmath}
\usepackage{amssymb}
\usepackage{enumitem}
\usepackage{xspace}
\usepackage{xcolor}
\usepackage{caption}
\usepackage{tcolorbox}
\tcbuselibrary{breakable,skins}
\newtcolorbox[auto counter]{caseprompt}[2][]{
  enhanced,breakable,boxrule=0.45pt,arc=1mm,
  colback=blue!2!white,colframe=blue!35!black,
  colbacktitle=blue!8!white,coltitle=black,
  fonttitle=\fontencoding{T1}\fontfamily{lmss}\fontseries{bx}\selectfont\small,
  fontupper=\fontencoding{T1}\fontfamily{lmtt}\selectfont\small,
  title={Prompt~\thetcbcounter: #2},
  title after break={Prompt~\thetcbcounter: #2 (continued)},
  left=8pt,right=8pt,top=6pt,bottom=6pt,
  before skip=9pt,after skip=9pt,
  before upper={\raggedright\setlength{\parskip}{5pt}},#1}

\title{Trace2Eval: Contextualized Evaluation of Coding Agents from Real-World Traces}

\author{%
\textbf{Ji Shi}$^{1}$ \quad \textbf{Lirong Jie}$^{1}$ \quad
\textbf{Lexiao Zou}$^{1}$ \quad \textbf{Yaoming Zhu}$^{2}$\\[3pt]
\textbf{Yifan Yang}$^{2}$ \quad \textbf{Xuezhi Cao}$^{2}$ \quad
\textbf{Xunliang Cai}$^{2}$ \quad \textbf{Miao Zhang}$^{1}$\\[5pt]
$^{1}$Harbin Institute of Technology, Shenzhen\\
$^{2}$Independent Researcher
}

\hypersetup{
  pdftitle={Trace2Eval: Contextualized Evaluation of Coding Agents from Real-World Traces},
  pdfauthor={Ji Shi, Lirong Jie, Lexiao Zou, Yaoming Zhu, Yifan Yang, Xuezhi Cao, Xunliang Cai, Miao Zhang},
  pdfsubject={Preprint},
  pdfkeywords={coding agents, evaluation, benchmarks, real-world traces}}

\newcommand{\method}{Trace2Eval\xspace}
\newcommand{\triple}{\ensuremath{\langle q, \widehat W_0, \widehat E_0\rangle}}

\iclrfinalcopy
\begin{document}
\ifdefined\XeTeXversion\fi
\maketitle
\lhead{Preprint}
\begin{abstract}
Coding agents are increasingly used in real-world workflows across diverse domains, motivating evaluation grounded in real user tasks. Existing benchmarks, however, largely rely on static public tasks with reference patches or hidden tests. Next-generation coding-agent benchmarks should therefore move beyond correctness on fixed tasks toward continually renewed evaluation of overall artifact quality in real user workflows.
We propose \method, a dynamically updatable pipeline that constructs benchmarks from real-world interaction trajectories. \method employs a multi-agent reconstruction system to analyze the filtered trajectory, recover the original user query, restore the workspace files to their pre-interaction state, and rebuild a trajectory-equivalent execution environment. A candidate agent then executes the reconstructed query to produce an initial artifact.
To evaluate these artifacts, the pipeline iteratively refines human rubrics through inter-annotator alignment, and calibrates judge agents against human assessments using source-code and runtime evidence. Guided by feedback from the \method pipeline, models revise their initial artifacts, yielding a mean score gain of 10.05~percentage points across 19 model-harness configurations and a maximum of 19.28~percentage points for DeepSeek-V4-Pro with Claude Code, which supports rubric feedback as actionable guidance for artifact improvement.\footnote{{Supporting materials are available at https://anonymous.4open.science/r/TRACE2EVAL-ICLR.}}
\end{abstract}

\section{Introduction}
\label{sec:introduction}
Powered by increasingly capable frontier models~\citep{openai_gpt56_2026,anthropic_fable5_2026}, coding agents such as Codex~\citep{openai_codex_2025} and Claude Code~\citep{anthropic_claude_2026} can inspect and edit repositories, execute commands, and complete tasks in users' workspaces. Benchmarks evolved from evaluating isolated code to assessing changes within existing software projects. HumanEval and MBPP assessed function-level solutions through tests~\citep{chen2021evaluating,austin2021program}. APPS introduced competition-style problems~\citep{hendrycks2021apps}, while DS-1000 and BigCodeBench expanded coverage to library-rich tasks~\citep{lai2023ds1000,zhuo2024bigcodebench}. LiveCodeBench added time-sliced problems to renew evaluation as models evolved~\citep{jain2024livecodebench}. SWE-bench moved to repository-level issue resolution, where agents worked from repository snapshots and produced patches assessed against test outcomes~\citep{jimenez2023swebench}.

Although this evolution brought evaluation closer to real development through more demanding repository-level tasks, important gaps remain between benchmark settings and real user workflows, raising several challenges for reliable evaluation. Public programming problems and repository issues make tasks and reference solutions increasingly retrievable, leaving static collections vulnerable to contamination and benchmark hacking~\citep{openai2026nolonger_swebench_verified,benchjack2026,magar2023contamination}. Time-sliced evaluation renewed task content~\citep{jain2024livecodebench}, but freshness alone does not establish whether the selected activities match evolving user demand: competition problems and issue-to-patch repair cover only part of the work users delegate to coding agents. Even for similar requests, a benchmark's prescribed repository snapshot may differ from the user's initial files and dependencies, changing the work needed to complete the task. Correctness tests then assess behavior encoded in their checks, while a delivered application can implement tested operations yet omit the controls needed to use them. Such an artifact may satisfy component-level checks without completing the user's workflow. Connecting scores to real use therefore requires representative requests, their original context, and criteria for the quality of the delivered artifact.

These observations motivate contextualized evaluation, which assesses the requests people delegate to coding agents within their working environments, including the initial files, dependencies, and delivery requirements. A sustainable benchmark for this setting should satisfy four properties at once. \textbf{(1) Authenticity.} Tasks should come from real user traffic, preserving the original goals and context that distinguish practical workflows from curated problems. \textbf{(2) Contamination resistance.} Evaluation should avoid reliance on publicly retrievable solutions, assessing delivered artifacts against user requirements rather than an available reference patch. \textbf{(3) Dynamism.} A refreshable construction procedure should periodically re-sample user traffic and re-score artifacts as users' goals, workflows, and agent capabilities evolve. \textbf{(4) Measurability.} Explicit scoring criteria should translate user-perceived quality into measurable judgments, making scores interpretable even when the original request is underspecified. No existing benchmark meets all four requirements. Meeting these requirements calls for a reusable evaluation protocol that follows evolving user needs while maintaining consistent standards for artifact quality.

\begin{figure}[t]
  \centering
  \includegraphics[width=\linewidth]{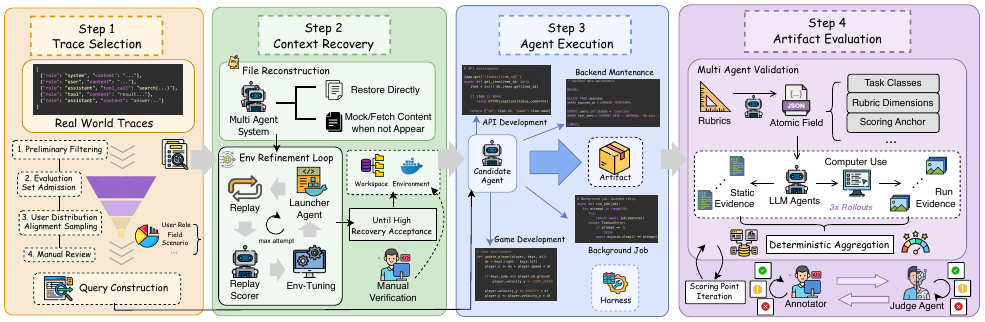}
  \caption{Overview of the \method workflow. \textbf{(1) Trace selection.} Two-stage greedy selection aligns candidates with online request profiles, then combines rubric-based admission with distribution constraints. \textbf{(2) Context recovery.} For each retained request, recovery agents reconstruct its initial workspace and environment, using trajectory replay to assess and refine the recovered context. \textbf{(3) Agent execution.} Candidate coding agents execute the request in this context and produce an artifact. \textbf{(4) Artifact evaluation.} Judge agents assess the artifact using source and runtime evidence and human-grounded rubrics. Repeated scoring yields dimension-level and aggregate quality scores.}
  \label{fig:overview}
\end{figure}

We propose \method, a reusable evaluation protocol implemented as a dynamic pipeline that constructs coding-agent benchmarks from real-world traces (Figure~\ref{fig:overview}). Given an anonymized interaction trace $\tau$ between a user and a coding agent, \method reconstructs a benchmark instance $x = \triple$, where $q$ is the user task distilled from the trace, $\widehat W_0$ is the recovered workspace before the recorded interaction, and $\widehat E_0$ is a containerized approximation of the user's initial execution environment. A candidate agent runs from $x$ to produce an artifact $y$, which judge agents assess against human-grounded rubrics. This accommodates open-ended tasks without requiring a unique reference patch. To further assess the rubric's validity, we translate diagnosed weaknesses into revision feedback and reassess the resulting artifacts. Verified repairs and score gains provide complementary evidence that the rubric captures meaningful aspects of user-perceived quality.

Our contributions are as follows:
\begin{itemize}[leftmargin=*, itemsep=2pt, topsep=2pt]
    \item \textbf{A dynamic, trace-driven evaluation paradigm.} We introduce evaluation-as-construction: rather than a static dataset, \method is a periodically executable pipeline that re-samples real online traffic and reconstructs $\triple$ instances, so the evaluation tracks both evolution of user intent and model capability.
    \item \textbf{Recovery of initial task contexts from incomplete traces.} We develop a reconstruction workflow that combines observed file evidence, contextual completion, and dependency inference to recover the initial workspace and environment. Bounded replay and refinement assess whether the recovered context supports the recorded tool behavior.
    \item \textbf{Rubric-based quantification of user experience.} We use rubrics to translate user-perceived delivery quality into explicit, dimension-specific scoring criteria, together with a standardized procedure for constructing ground truth and aligning judge agents with human evaluations.
    \item \textbf{Evaluation and rubric-guided artifact improvement.} To test whether rubric-based assessment provides useful improvement signals, we evaluate initial artifacts, turn diagnosed weaknesses into revision instructions, and reassess the revised artifacts. Total scores rise by 10.05~pp on average, with DeepSeek-V4-Pro gaining 19.28~pp under Claude Code (Table~\ref{tab:rq3-panel}), providing complementary evidence for the practical value of our rubric and scoring framework.
\end{itemize}

\section{Related Work}
\label{sec:related}
\paragraph{Static code and repository-level evaluation.}
Repository-level evaluation expanded from issue resolution in SWE-bench and its extensions~\citep{jimenez2023swebench,openai2024sweverified,deng2025swebenchpro} to project-level code generation in DevEval~\citep{li2024deveval} and repository construction in NL2Repo-Bench~\citep{nl2repobench2025}. Although this progression broadened task scope, predefined tasks and correctness criteria still capture only part of what users expect from coding agents. Public benchmark exposure further threatens evaluation validity~\citep{openai2026nolonger_swebench_verified}, while changing user goals make fixed task distributions less representative over time. These limitations motivate \method's use of fresh user traces, which preserve the context of actual requests and support evaluation of delivered artifacts as usage evolves.

\paragraph{Environment construction.}
Environment construction supports scalable task generation through several routes. SWE-smith created test-breaking tasks in repository environments~\citep{yang2025swesmith}, whereas SWE-Mirror transferred issue semantics into repositories with configured environments~\citep{swemirror2025}. These approaches expanded task collections by exploiting existing code and tests, reducing the need to configure each task independently. For terminal tasks, TerminalTraj and TermiGen constructed Docker environments for trajectory synthesis~\citep{terminaltraj2026,termigen2026}, while CLI-Universe combined taxonomy-guided task generation with executable verification~\citep{cliuniverse2026}. TerminalWorld reconstructed terminal tasks from public recordings and evaluated completion with executable tests~\citep{terminalworld2026}. \method instead recovers the initial contexts of user--agent interactions and assesses delivered coding artifacts across functionality, code quality, and user experience with human-grounded rubrics.

\paragraph{Rubrics, judges, and feedback.}
Open-ended evaluation combines explicit quality criteria with judgments whose reliability must be established. HELM advocated measuring multiple dimensions~\citep{liang2022helm}, while rubric research studied how such dimensions become criteria for evaluation and training~\citep{rubricsurvey2025}. G-Eval used structured criteria to improve agreement with human assessments~\citep{li2023geval}; MT-Bench examined both the agreement and biases of LLM judges~\citep{li2023mtbench}, and AlpacaEval provided a human-validated automatic evaluator~\citep{li2023alpacaeval}. Together, these studies motivate checking automated scores against human judgments rather than assuming that explicit criteria alone ensure reliability. Beyond scoring, Self-Refine used feedback to revise generated outputs~\citep{madaan2023selfrefine}; Asuka-Bench and CATArena studied feedback-driven development and iterative code improvement~\citep{asuka2026,catarena2025}, connecting assessment with improvement. \method uses judge agents to automatically score coding artifacts against human-grounded rubrics, drawing on source-code and runtime evidence.

\section{\method}
\label{sec:method}
Starting from anonymized real-world user traces, \method builds executable evaluation instances for coding agents and scores their artifacts with multi-dimensional rubrics rather than a single reference patch. As shown in Figure~\ref{fig:overview}, the pipeline first retrieves and labels user traces (\S\ref{sec:trace-retrieval}), then recovers an evaluation instance \triple, including the user query $q$, recovered workspace $\widehat W_0$, and recovered execution environment $\widehat E_0$ (\S\ref{sec:recovery}). The evaluated model is then run from the recovered instance to produce an artifact, which is assessed with human-grounded rubrics and a static-dynamic LLM-as-judge framework (\S\ref{sec:rubric}). This design evaluates agents in the original user context rather than a curated GitHub-issue setting. By periodically selecting new online traces and updating the dataset, \method keeps evaluation aligned with evolving real-world usage.
\subsection{Retrieving, Filtering, and Labeling Online Traces}
\label{sec:trace-retrieval}

We collect 20,000 anonymized traces from online coding-agent traffic between June and July 2026 by randomly sampling approximately equal numbers per day, of which 14,603 are identified as code-related and form the population for our distribution analysis. We label coding traces using the 171-category, six-axis taxonomy summarized in Appendix~\ref{app:full-profile-tables}. Its four top-level dimensions are task type (\texttt{x1}/\texttt{x2}), usage scenario (\texttt{y}), technical domain (\texttt{z1}/\texttt{z2}), and user role (\texttt{w}). A two-stage greedy procedure first selects 300 distribution-aligned candidates (Algorithm~\ref{alg:align-select}), then combines admission-rubric voting with distribution constraints to select the final 129 tasks (Table~\ref{tab:admission-rules}).

\begin{figure}[t]
  \centering
  \includegraphics[width=\linewidth]{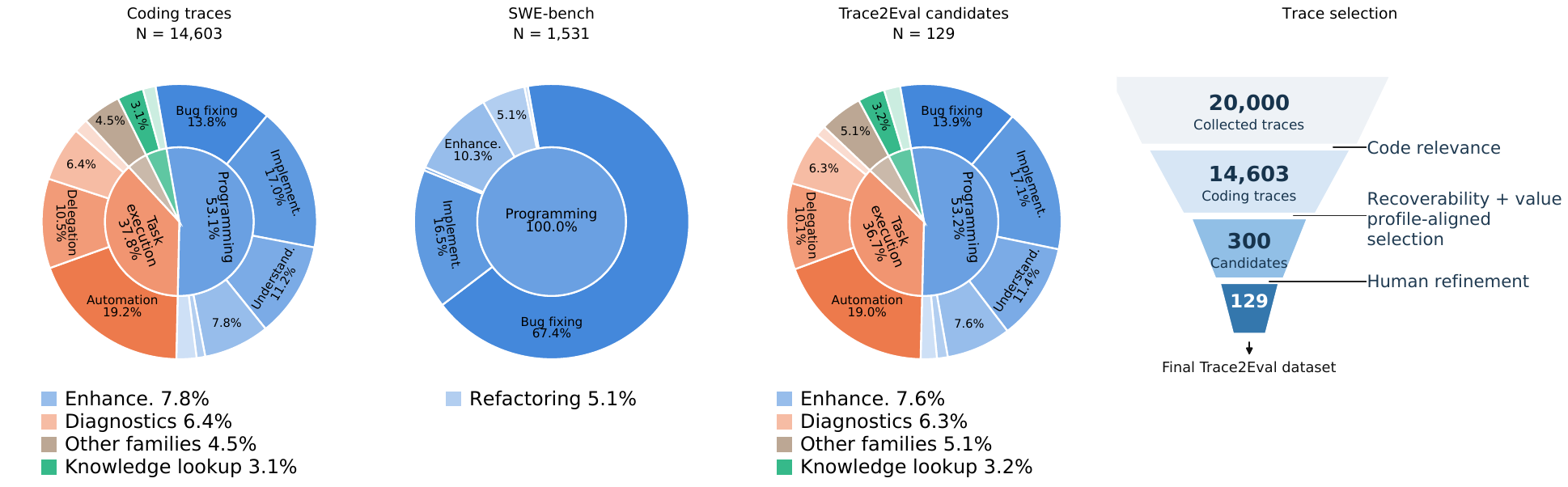}
  \caption{Task-type distributions and trace selection. The pies compare online coding traces, SWE-bench, and selected candidates; inner disks show broad task families (\texttt{x1}), and outer rings show finer activities (\texttt{x2}). Candidates retain the broader mix of online requests, whereas SWE-bench concentrates on programming and bug fixing. The funnel shows the successive selection stages leading to the final benchmark.}
  \label{fig:profiles-task}
\end{figure}

Figure~\ref{fig:profiles-task} compares the task-type distributions of online coding traces, SWE-bench, and the final 129-task \method benchmark, alongside the selection stages. The visualization shows that the selected candidates cover common activities in online coding requests, whereas SWE-bench concentrates more heavily on debugging. This broader coverage connects evaluation to the range of work users delegate to coding agents. Appendix~\ref{app:data-construction} extends the comparison across all six profile sub-axes. The second greedy stage preserves distribution alignment while admitting tasks that meaningfully distinguish coding-agent capabilities (Appendix~\ref{app:selection-algorithm}).

\subsection{Workspace and Environment Recovery}
\label{sec:recovery}

Online traces provide tool observations rather than complete snapshots of the user's workspace and environment. A complete ground-truth state is therefore unavailable for direct comparison, so we assess recovery through the behavior of the reconstructed context. We use \emph{trajectory equivalence} as the joint acceptance criterion for the recovered workspace and environment: whether replaying the original tool calls reproduces their recorded results.

\paragraph{Workspace and environment recovery.}
Let $W_0$ and $E_0$ denote the original initial workspace and environment, with hats marking recovered estimates. We reconstruct $\widehat W_0$ using file evidence from the trace, restoring directly observed content and filling gaps through contextual inference or retrieval from verified public sources. The recovered workspace $\widehat W_0$ excludes artifacts introduced during the session and restores modified pre-existing files to their initial state. Environment recovery similarly separates pre-existing dependencies from changes made during execution. Let $\tau=(T_1,\ldots,T_T)$ be a trace of $T$ tool calls, with $T_t=(u_t,i_t,o_t)$ denoting the tool name, input, and recorded output at step $t$. We infer the initial environment $E_0$ and each step's environment change $\Delta E_t$ through the update
\begin{equation}
    \label{eq:env-step}
    \operatorname{Agent}\bigl(C_{E_t}, C_{E_0}^{(t)}\mid T_t\bigr)
    =\bigl(C_{E_{t+1}}, C_{E_0}^{(t+1)}\bigr),
\end{equation}
where $\operatorname{Agent}$ is the recovery agent operating on two container states: $C_{E_t}$ is the current replay environment before call $t$, while $C_{E_0}^{(t)}$ is the evolving estimate of the initial environment after the preceding calls. Within each replay round, call outcomes guide updates to both states, distinguishing initial dependencies from later changes. The resulting initial-state container realizes $\widehat E_0$, while the inferred stepwise changes define $\widehat{\Delta E}_t$. Multiple replay rounds refine both states before we package $\widehat E_0$ as the evaluation image. Appendix~\ref{app:recovery} details the file-recovery agents and environment updates.

\paragraph{Trajectory equivalence through replay.}
We extract tool calls from the original trace and replay them in order in the recovered workspace and cold-started environment. For each call, a replay evaluation agent compares the replay result $a_t$ with the recorded result $o_t$, using the tool and input as context, and assigns
\begin{equation}
    m_t=\operatorname{Judge}_{\mathrm{replay}}(u_t,i_t;a_t,o_t)\in[0,1].
\end{equation}
The score $m_t$ comes from the replay evaluation agent's judgment of agreement: it assigns 1 when the results are identical and lower scores for more substantial differences. For each replay round, let $\sI(\tau)$ contain the indices of calls with scored comparisons. For $|\sI(\tau)|>0$, the \emph{replay score} quantifies trajectory equivalence as the arithmetic mean of these per-call scores:
\begin{equation}
    \label{eq:equiv-main}
    \operatorname{ReplayScore}(\widehat W_0,\widehat E_0;\tau)=\frac{1}{|\sI(\tau)|}\sum_{t\in\sI(\tau)}m_t.
\end{equation}
When all tool calls in $\tau$ are replayed and scored, a replay score of 1 indicates that the recovered workspace and environment reproduce the original results throughout the trajectory. This trajectory-equivalent workspace and environment form our recovery target, with the replay score providing the quantitative basis for recovery acceptance. Recovery agents repair the file or environment defects revealed by low-scoring calls, then replay the trajectory from the revised context to verify the corrections. This cycle continues until the discrepancies are resolved or the replay budget is reached, followed by human verification (Appendix~\ref{app:recovery}).

\subsection{Rubric Construction Protocol and LLM-as-Judge Scoring}
\label{sec:rubric}
We assess artifacts produced in the recovered task contexts using a human-grounded rubric, combining expert design with calibrated agent judgment.

\paragraph{Expert-guided rubric construction.}
A qualitative analysis of human preference rationales identifies the quality concerns that inform the rubric (Appendix~\ref{app:rubric-basis}). Experts review their relevance and scope, confirm 11 dimensions across accuracy, code quality, and user experience, and define initial anchors for scores $h_{i,d}\in\{0,1,2\}$ for artifact $i$ on dimension $d$. They set group weights of 50\%, 35\%, and 15\%, respectively, and within-group priorities based on expert judgment of their importance for user-perceived quality (Table~\ref{tab:rubric-dims}). This specification provides the starting point for subsequent human and agent calibration.

\paragraph{Static-Dynamic LLM-as-judge scoring.}
Each rubric dimension specifies the evidence required for scoring. Seven dimensions are assessed from source code, while code execution, data authenticity, interface and interaction, and information presentation require runtime or rendered evidence. Static analysis and runtime agents collect evidence for the judge agent, implemented by GLM-5.3-Flash in the model comparison. For each instance $i$ and dimension $d$, the scoring workflow performs $K$ independent evidence-collection rollouts, indexed by $k$. Let $e^{(k)}_{i,d}$ be the evidence collected in rollout $k$, and let $R_d$ be the rubric anchor for dimension $d$. The judge agent $J$ maps evidence to a discrete rubric score,
\begin{equation}
    \hat{h}^{(k)}_{i,d}\;=\;J\bigl(R_d,\,e^{(k)}_{i,d}\bigr)\;\in\;\{0,1,2\}.
\end{equation}
A deterministic aggregator combines the $K$ rollout scores into the final score, consensus label, and self-consistency statistics (Appendix~\ref{app:scoring}).

\paragraph{Rubric-level ground truth.}
Since we need a reference score for each artifact to iteratively refine the rubric definitions and score anchors, we define rubric-level ground truth as the adjudicated human score for each artifact--dimension pair, denoted by $g_{i,d}\in[0,2]$. The first two human annotation rounds provide independent scores and evidence-based rationales for each artifact. In the third round, a third annotator reviews both sets of scores and rationales alongside the artifact itself, then assigns the final reference score $g_{i,d}$ (Appendix~\ref{app:annotation}).

Calibration proceeds in two stages. We first refine the human score anchors to achieve high human-human agreement and establish the adjudicated reference $g_{i,d}$. We then give the judge agent this human rubric as its initial scoring instructions. Because its scores do not align sufficiently with the human reference, we refine the agent scoring points over multiple rounds while keeping the dimensions, weights, and human reference fixed. The resulting agent anchors specify evidence requirements and decision boundaries learned from these disagreements. With these refinements, judge agents reach human-level agreement on several dimensions, supporting scalable rubric-based evaluation (detailed in Appendices~\ref{app:rubric-scoring} and \ref{app:judge-calibration}).



\section{Experiments}
\label{sec:experiments}
We organize the evaluation around three research questions:
\begin{itemize}[leftmargin=*,itemsep=1pt,topsep=2pt,parsep=0pt]
    \item \textbf{RQ1:} How faithfully can \method recover the initial workspace and environment from user traces?
    \item \textbf{RQ2:} How accurate is the evaluation? Does the rubric reflect user-perceived quality (RQ2.1), and do judge agents agree with human ground truth (RQ2.2)?
    \item \textbf{RQ3:} Can the evaluation guide agent improvement? What capabilities do models exhibit across harnesses (RQ3.1), and does rubric-derived feedback improve their artifacts (RQ3.2)?
\end{itemize}

\subsection{Experimental Setup}
\label{sec:setup}
Candidate models with specific harness first execute the 129 \method tasks in their recovered contexts through Claude Code or OpenHands within Terminal-Bench~\citep{terminalbench2026}. GLM-5.3-Flash evaluates the resulting artifacts through three scoring rollouts under the calibrated rubric. GLM-5.2 diagnosis, design, and review agents turn rubric-derived feedback into revision queries; each model then improves its artifact using the same harness and preserved task environment. Re-evaluation under the same rubric tests whether the assessment provides useful guidance for improvement. Appendix~\ref{app:experiment-plan} documents execution and component validation, while Appendix~\ref{app:feedback} details feedback preparation and revision.

\begin{table}[!t]
\centering\scriptsize
\setlength{\tabcolsep}{2pt}
\caption{Artifact quality before (R1) and after rubric-guided revision (R2) on the full 129-task \method benchmark. Scores use \% of maximum; changes are in percentage points (pp). Acc, Code, and UX denote accuracy, code quality, and user experience. Within each harness and column, bold and underlining mark the best and second-best values. Model names link to official documentation, cited in Appendix~\ref{app:model-panel-configuration}.
}
\label{tab:rq3-panel}
\resizebox{\linewidth}{!}{%
\begin{tabular}{l*{8}{r}@{\hspace{7pt}}*{4}{r}}
\toprule
\multirow{2}{*}{Model} & \multicolumn{2}{c}{Acc} & \multicolumn{2}{c}{Code} & \multicolumn{2}{c}{UX} & \multicolumn{2}{c}{Total} & \multicolumn{4}{c}{Change (pp)} \\
\cmidrule(lr){2-3}\cmidrule(lr){4-5}\cmidrule(lr){6-7}\cmidrule(lr){8-9}\cmidrule(lr){10-13}
 & R1 & R2 & R1 & R2 & R1 & R2 & R1 & R2 & $\Delta$Acc & $\Delta$Code & $\Delta$UX & $\Delta$Total \\
\midrule
\multicolumn{13}{l}{\textbf{Claude Code}} \\
\href{https://huggingface.co/deepseek-ai/DeepSeek-V4-Pro}{DeepSeek-V4-Pro} & 75.10 & \underline{97.19} & 82.40 & 93.14 & 66.30 & 89.50 & 76.85 & 96.13 & \textbf{+22.09} & \underline{+10.74} & +23.20 & \textbf{+19.28} \\
\href{https://huggingface.co/zai-org/GLM-5.2}{GLM-5.2} & 79.80 & 94.80 & 86.90 & 96.46 & 68.40 & 88.01 & 80.96 & 94.13 & +15.00 & +9.56 & +19.61 & +13.17 \\
\href{https://huggingface.co/zai-org/GLM-5.3}{GLM-5.3} & 83.50 & 97.13 & 90.70 & \underline{99.18} & 70.10 & \textbf{93.43} & 84.76 & \underline{97.29} & +13.63 & +8.48 & \textbf{+23.33} & +12.53 \\
\href{https://developers.openai.com/api/docs/models/gpt-5.5}{GPT-5.5} & 76.80 & 89.04 & 86.70 & 95.83 & 59.40 & 78.24 & 78.10 & 89.88 & +12.24 & +9.13 & +18.84 & +11.78 \\
\href{https://developers.openai.com/api/docs/models/gpt-5.6-sol}{GPT-5.6 sol} & \underline{85.93} & 94.37 & \underline{92.44} & 97.36 & \underline{74.82} & 88.47 & \underline{86.23} & 94.87 & +8.44 & +4.92 & +13.65 & +8.64 \\
\href{https://huggingface.co/moonshotai/Kimi-K3}{Kimi-K3} & 78.43 & 88.27 & 87.92 & 93.86 & 60.24 & 76.49 & 78.93 & 88.07 & +9.84 & +5.94 & +16.25 & +9.14 \\
\href{https://huggingface.co/MiniMaxAI/MiniMax-M3}{MiniMax-M3} & 75.80 & 90.85 & 85.40 & 96.24 & 58.90 & 82.20 & 76.64 & 91.57 & \underline{+15.05} & \textbf{+10.84} & \underline{+23.30} & \underline{+14.93} \\
\href{https://huggingface.co/meituan-longcat/LongCat-2.0}{LongCat-2.0} & 82.20 & 90.94 & 90.85 & 96.13 & 64.40 & 79.81 & 82.65 & 92.04 & +8.74 & +5.28 & +15.41 & +9.39 \\
\href{https://www.anthropic.com/news/claude-fable-5-mythos-5}{Fable 5} & \textbf{92.31} & \textbf{97.26} & \textbf{96.05} & \textbf{99.28} & \textbf{81.76} & \underline{90.93} & \textbf{92.42} & \textbf{97.37} & +4.95 & +3.23 & +9.17 & +4.95 \\
\midrule
\multicolumn{13}{l}{\textbf{OpenHands}} \\
\href{https://huggingface.co/deepseek-ai/DeepSeek-V4-Pro}{DeepSeek-V4-Pro} & 87.85 & 92.13 & 92.01 & 97.59 & 75.00 & 89.89 & 85.72 & 93.75 & +4.28 & +5.58 & +14.89 & +8.03 \\
\href{https://ai.google.dev/gemini-api/docs/models/gemini-3.6-flash}{Gemini 3.6 Flash} & 82.56 & 91.02 & 90.15 & 96.10 & 70.68 & 86.24 & 82.89 & 91.76 & +8.46 & \underline{+5.95} & +15.56 & +8.87 \\
\href{https://huggingface.co/zai-org/GLM-5.2}{GLM-5.2} & 86.90 & 94.24 & 93.03 & \textbf{98.67} & 67.52 & 88.53 & 84.27 & 95.09 & +7.34 & +5.64 & \underline{+21.01} & \underline{+10.82} \\
\href{https://huggingface.co/zai-org/GLM-5.3}{GLM-5.3} & 82.10 & 90.43 & 90.19 & 95.10 & 72.07 & 78.55 & 84.90 & 90.41 & +8.33 & +4.91 & +6.48 & +5.51 \\
\href{https://developers.openai.com/api/docs/models/gpt-5.5}{GPT-5.5} & 84.11 & 93.30 & 90.13 & 96.03 & 65.47 & 88.33 & 82.86 & 94.62 & +9.19 & +5.90 & \textbf{+22.86} & \textbf{+11.76} \\
\href{https://developers.openai.com/api/docs/models/gpt-5.6-sol}{GPT-5.6 sol} & \underline{88.23} & \textbf{95.36} & \textbf{93.54} & \underline{98.33} & \textbf{78.14} & \textbf{93.38} & \underline{87.46} & \underline{95.89} & +7.13 & +4.79 & +15.24 & +8.43 \\
\href{https://huggingface.co/moonshotai/Kimi-K3}{Kimi-K3} & 77.04 & 86.67 & 86.84 & 92.76 & 59.43 & 76.27 & 77.54 & 86.16 & \textbf{+9.63} & +5.92 & +16.84 & +8.62 \\
\href{https://huggingface.co/MiniMaxAI/MiniMax-M3}{MiniMax-M3} & 84.27 & 90.08 & 90.67 & 98.05 & 65.90 & 86.07 & 82.40 & 91.97 & +5.81 & \textbf{+7.38} & +20.17 & +9.57 \\
\href{https://huggingface.co/meituan-longcat/LongCat-2.0}{LongCat-2.0} & 78.72 & 88.18 & 88.21 & 93.96 & 61.43 & 78.39 & 79.52 & 88.18 & \underline{+9.46} & +5.75 & +16.96 & +8.66 \\
\href{https://www.anthropic.com/news/claude-fable-5-mythos-5}{Fable 5} & \textbf{88.87} & \underline{95.12} & \underline{93.06} & 97.58 & \underline{76.57} & \underline{93.11} & \textbf{88.98} & \textbf{95.91} & +6.25 & +4.52 & +16.54 & +6.93 \\
\bottomrule
\end{tabular}%
}
\end{table}

\subsection{Recovery Fidelity (RQ1)}
\label{sec:rq1}
Each instance \triple{} uses a retained user request $q$, with initial workspace $\widehat W_0$ and environment $\widehat E_0$ reconstructed from file and dependency evidence in its trace. To validate the usability of the recovered workspace and environment, human reviewers inspect the initial files for completeness, while replay tests tool behavior in the cold-started environment. The replay evaluation agent compares tool-call results with the original observations (\S\ref{sec:recovery}).

On the recovery calibration set, we compare replay fidelity before and after refinement. The difficulty-stratified analysis shows larger gains where startup repair leaves behavioral discrepancies, indicating that replay identifies defects beyond those exposed at launch. Appendices~\ref{app:recovery-calibration}--\ref{app:environment-calibration} detail the comparisons. With an 82.6\% workspace acceptance rate and a mean replay score of 0.9175, these results answer RQ1 by demonstrating that recovery can provide high-quality initial workspaces and environments for executable agent evaluation grounded in real users' development contexts.

\subsection{Evaluation Accuracy (RQ2)}
\label{sec:rq2}
\textbf{Alignment with user-perceived quality (RQ2.1).} To ground artifact evaluation in real development needs, we collect expert judgments of the relative importance of the quality concerns identified in human preference rationales (Appendix~\ref{app:rubric-basis}). These judgments determine the final group weights and within-group priorities in Table~\ref{tab:rubric-dims}.

For example, the final weights of 22.5\% for requirement implementation and 20.0\% for code execution preserve the experts' relative priorities. This alignment emphasizes completing the user's request and delivering a runnable artifact, reflecting expert perceptions of user needs and addressing RQ2.1.

\textbf{Human agreement (RQ2.2).} To test whether this human-grounded rubric supports automated evaluation, we compare judge agents with adjudicated human ground truth (GT; \S\ref{sec:rubric}). Figure~\ref{fig:judge-calibration}, Table~\ref{tab:rq2-kappa}, and Appendix~\ref{app:judge-calibration} provide the criterion-level evidence. The comparisons distinguish usability agreement, based on zero versus nonzero grades, from exact agreement across quality levels. Cohen\textquotesingle{}s $\kappa=(p_{\mathrm{o}}-p_{\mathrm{e}})/(1-p_{\mathrm{e}})$ adjusts observed agreement $p_{\mathrm{o}}$ for agreement $p_{\mathrm{e}}$ expected by chance from the grade distributions.

The judge agent achieves human-comparable or higher exact agreement on selected dimensions, such as architecture. Feature richness also shows an advantage over the human average in both exact agreement and $\kappa$. Overall usability agreement of 93.0\%, compared with 92.9\% for humans, validates the judge agent\textquotesingle{}s accuracy in identifying artifact usability and supports its use as a partial substitute for human assessment under the rubric, answering RQ2.2.

\subsection{Agent Performance and Feedback (RQ3)}
\label{sec:rq3}
We compare initial delivery and feedback-guided revision to examine model capabilities and validate the rubric's guidance. R1 denotes initial execution; R2 revises the resulting artifact using rubric feedback, with the same model, harness, and preserved environment.

\begin{figure}[t]
\centering
\begin{minipage}[c]{.46\linewidth}
\centering
\captionsetup{font=normalsize,skip=4pt,hypcap=false}
\includegraphics[width=\linewidth,height=4in,keepaspectratio]{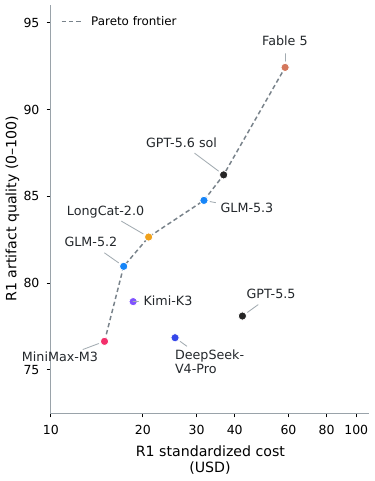}
\caption{R1 artifact quality versus standardized execution cost under Claude Code. The dashed line connects Pareto-efficient models.}
\label{fig:cost-effectiveness}
\end{minipage}\hfill
\begin{minipage}[c]{.53\linewidth}
\centering
\captionsetup{font=normalsize,skip=4pt,hypcap=false}
\includegraphics[width=\linewidth]{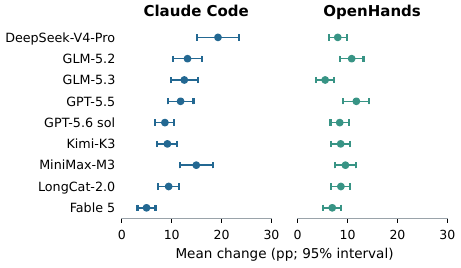}
\caption{Mean R2--R1 Total gains (pp) under each harness. Bars show paired-task bootstrap 95\% CI.}
\label{fig:feedback-gains}

\vspace{6pt}
\captionof{table}{Task-level outcomes on all 129 Trace2Eval tasks. Imp., Same, and Reg. denote increased, unchanged, and decreased Total scores.}
\label{tab:feedback-task-outcomes}
\begingroup
\footnotesize
\setlength{\tabcolsep}{0.8pt}
\renewcommand{\arraystretch}{1.03}
\begin{tabular*}{\linewidth}{@{\extracolsep{\fill}}lrrrrrr@{}}
\toprule
& \multicolumn{3}{c}{Claude Code} & \multicolumn{3}{c}{OpenHands} \\
\cmidrule(lr){2-4}\cmidrule(l){5-7}
Model & Imp.$\uparrow$ & Same$\downarrow$ & Reg.$\downarrow$ & Imp.$\uparrow$ & Same$\downarrow$ & Reg.$\downarrow$ \\
\midrule
DeepSeek-V4-Pro & 70 & 28 & 31 & 85 & 23 & 21 \\
GLM-5.2 & 84 & 15 & 30 & $\underline{106}$ & $\underline{11}$ & $\underline{12}$ \\
GLM-5.3 & 83 & 27 & \textbf{19} & 94 & \textbf{10} & 25 \\
GPT-5.5 & \textbf{95} & \textbf{11} & $\underline{23}$ & \textbf{108} & 12 & \textbf{9} \\
GPT-5.6 sol & 83 & $\underline{13}$ & 33 & 91 & $\underline{11}$ & 27 \\
Kimi-K3 & 84 & $\underline{13}$ & 32 & 88 & $\underline{11}$ & 30 \\
MiniMax-M3 & $\underline{89}$ & 15 & 25 & 97 & 14 & 18 \\
LongCat-2.0 & 88 & $\underline{13}$ & 28 & 87 & 12 & 30 \\
Fable 5 & 71 & 24 & 34 & 86 & 13 & 30 \\
\bottomrule
\end{tabular*}
\endgroup

\end{minipage}
\end{figure}

\paragraph{Model capability profiles (RQ3.1).}
We first examine whether stronger code quality also yields better user experience in R1 (Table~\ref{tab:rq3-panel}). Under OpenHands, GLM-5.2 leads DeepSeek-V4-Pro in code quality but trails it by 7.48~pp in user experience and also has a lower Total score. Under Claude Code, LongCat-2.0 and GLM-5.3 have nearly identical code-quality scores (90.85 versus 90.70), yet GLM-5.3 leads by 5.70~pp in user experience. Figure~\ref{fig:cost-effectiveness} adds execution cost to these profiles. Across both harnesses, these comparisons show why code quality alone cannot characterize practical delivery: \method distinguishes models by how well their artifacts serve users, answering RQ3.1.

\paragraph{Feedback reveals different model strengths (RQ3.2).}
Beyond agreement with human judgments, we test whether the rubric's diagnoses help models improve their artifacts. Under Claude Code, DeepSeek-V4-Pro shows strong repair ability despite weak initial delivery: its Total rises from 76.85 to 96.13 (+19.28~pp), the largest gain in Table~\ref{tab:rq3-panel}. This response indicates that initial task completion alone does not capture how effectively a model can use explicit quality feedback. Fable 5 presents a complementary strength: it delivers the strongest initial artifacts (92.42) and refines them to 97.37, nearing the scoring ceiling. Feedback use also depends on the harness: GLM-5.3 starts at nearly identical Total scores under Claude Code and OpenHands (84.76 versus 84.90), but finishes 6.88~pp higher under Claude Code after revision. Rubric feedback thus supports both substantial repair and further refinement of already strong artifacts, answering RQ3.2 (Figure~\ref{fig:feedback-gains}; Table~\ref{tab:feedback-task-outcomes}).

\begin{figure}[!htbp]
\centering
\includegraphics[width=\linewidth]{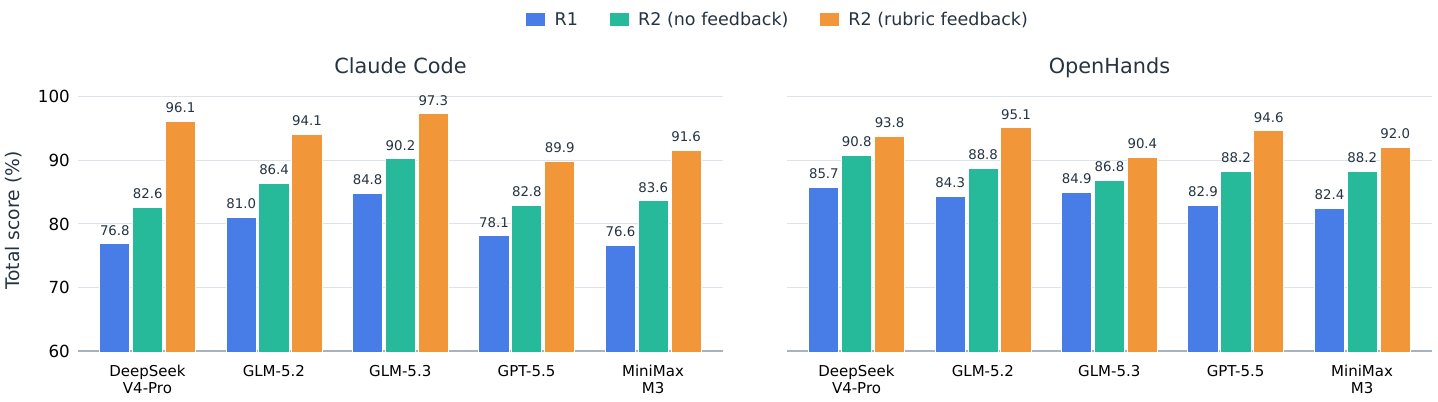}
\caption{Total scores for five representative models under Claude Code and OpenHands. Both revisions begin with the same R1 artifact: one uses only the original request, while the other receives rubric feedback. The bars compare R1, unguided R2, and rubric-guided R2 using complete experimental results, with rubric-guided scores also reported in Table~\ref{tab:rq3-panel}.}
\label{fig:feedback-ablation}
\end{figure}

\paragraph{Feedback ablation.}
To determine whether R2 improves because of rubric feedback or another development opportunity, we construct an unguided condition for each model. It starts from the model's R1 artifact and environment and repeats the original prompt unchanged. The guided condition uses the same starting state with rubric-derived repair instructions. Figure~\ref{fig:feedback-ablation} and Table~\ref{tab:feedback-ablation} compare both against R1. The smaller unguided gains suggest that another development pass helps, but targeted diagnoses provide substantial additional benefit. This comparison supports the answer to RQ3.2 by linking rubric feedback to artifact improvement (Appendix~\ref{app:feedback-ablation}).

\paragraph{Case studies: Feedback-guided artifact repair.}
We inspect concrete repairs to test whether score gains correspond to more usable workflows. In a workflow-management task, DeepSeek-V4-Pro initially implements backend APIs and client-side calls but leaves a placeholder frontend. Rubric feedback identifies the missing interface; the model adds pages, forms, and navigation. Browser evidence verifies these interactions, and interaction and information-presentation grades both rise from 0 to 2. This repair turns implemented components into functionality users can operate (Appendix~\ref{app:feedback-qualitative}). Appendix~\ref{app:feedback-case-study} traces a complementary repair from a failed build to a working login service.

The MiniMax-M3 commerce case (Appendix~\ref{app:feedback-qualitative}) illustrates incomplete repair: feedback leads to new backend endpoints, but inconsistent frontend routes and requests still interrupt the workflow, leaving code execution at grade~1. Gains thus depend on how fully the model carries a diagnosed repair through related components. Checking end-to-end usability lets the rubric distinguish partial edits from completed repairs.

The model-specific improvements and verified workflow repairs show that \method provides actionable feedback while distinguishing initial delivery, responsiveness to diagnosis, and the ability to complete a repair. This supports the rubric's practical validity and answers RQ3.2.

\section{Discussion}
\label{sec:discussion}
\paragraph{Contextualized evaluation and actionable feedback.}
Recovery enables evaluation in contexts drawn from real requests, while replay validates the recorded behavior (\S\ref{sec:rq1}). Two-stage greedy selection combines distribution alignment with rubric-based admission to retain meaningful evaluation tasks (Appendix~\ref{app:trace-retrieval}). Within these contexts, code quality and user experience reveal complementary aspects of delivery. The feedback ablation links dimension-level diagnosis to artifact improvement beyond an unguided revision (\S\ref{sec:rq3}), showing how evaluation can also guide development. The qualitative cases suggest that guidance is most useful when it identifies a concrete defect and a verifiable repair. Benefits depend on whether the candidate carries that repair through the affected components, explaining the coexistence of average gains and task-level regressions.

\paragraph{Limitations.}
Although judge agents achieve agreement comparable to human annotators on several dimensions, human--agent agreement still leaves room for improvement. We are refining the judge agents to improve consistency with human assessments. In addition, the file and environment recovery workflow occasionally produces task packages that lack core files even after human verification. Such omissions can obstruct subsequent agent execution, highlighting the need for further checks on the completeness of recovered workspaces.

\section{Conclusion}
\label{sec:conclusion}
We present \method, a refreshable procedure that turns real coding-agent traces into contextualized evaluation tasks. By recovering the initial workspace and environment, validating observed behavior through replay, and assessing artifacts with human-grounded rubrics, it connects the circumstances of a user request to explicit criteria for delivery quality. The component studies establish the feasibility and current limits of this construction, while evaluation on the 129-task benchmark reveals differences between code quality and user experience that motivate dimension-level assessment. Together, they position trace-driven construction as a basis for evaluation that can be renewed as user needs evolve, with each version's coverage and measurement reliability assessed explicitly.

\section*{AI use statement}
Generative AI supports three parts of this work. For \emph{benchmark construction}, Claude Sonnet 4.6 labels traces and votes on admission. Claude Opus 4.6 powers the restorer, mocker, fetcher, and validator agents for file recovery, together with the environment recovery, tuning, launcher, and replay evaluation agents (Appendices~\ref{app:data-construction} and~\ref{app:recovery}). For \emph{evaluation and revision}, GLM-5.3-Flash serves as the rubric-scoring judge agent for model comparison, using source and runtime evidence. GLM-5.2 performs the diagnosis, design, and review roles that prepare revision instructions (Appendices~\ref{app:rubric-scoring} and~\ref{app:feedback}). Candidate models execute tasks through Claude Code or OpenHands. For \emph{implementation and writing}, Codex and Claude Code assist workflow development, and AI tools help draft and polish the manuscript. AI-assisted contributions undergo human review, and the authors take responsibility for the final content.

\section*{Ethics statement}
\method draws on real histories of online coding-agent use. When using the service, users are informed that their data may be used for research after desensitization and provide explicit consent through signed agreements or informed-consent forms, conditional on protection of their personal privacy. The research data come from these consenting users and are desensitized before use. The public supplementary materials contain methodological documentation, rubric specifications, profile summaries, and selected case prompts; the full source-traffic collection is not included. These materials support the study of task completion and artifact quality and should not be used to identify or profile the original users. Reuse must respect confidentiality and applicable data and software licences. Human reference judgments support oversight of automated scoring (Appendix~\ref{app:rubric-scoring}), whose agreement varies across quality dimensions. The benchmark reflects a specific traffic population and expert-defined criteria, so its scores characterize artifact quality in that setting rather than universal user preferences or individual developer ability.

\section*{Reproducibility statement}
The accompanying \href{https://anonymous.4open.science/r/TRACE2EVAL-ICLR}{anonymous repository} provides workflow documentation, profile taxonomy and distribution tables, complete human and agent scoring anchors, and the original and feedback prompts for the revision case study. Section~\ref{sec:method} defines the construction and scoring methods, with trace selection and context recovery detailed in Appendices~\ref{app:data-construction} and~\ref{app:recovery}. Appendix~\ref{app:rubric-scoring} specifies dimension weights, human-reference calibration, repeated scoring, and aggregation, while Section~\ref{sec:experiments} and Appendix~\ref{app:experiment-plan} document experimental configurations and component validation. Appendix~\ref{app:feedback} describes revision inputs, preserved task states, task pairing, and uncertainty estimation, alongside analysis of recorded repairs; Appendix~\ref{app:feedback-ablation} specifies the unguided comparison. These descriptions connect the workflow to its validation criteria. Machine-readable taxonomy and rubric files accompany the human-readable specifications in the repository, supporting inspection and reuse of the evaluation protocol.

\begingroup
\RaggedRight
\bibliography{iclr2026_conference}
\bibliographystyle{iclr2027_conference}
\endgroup
\newpage
\appendix
\raggedbottom
\section{Supplementary Methods and Results}
\label{sec:appendix}

The appendix expands the construction and evaluation protocols in the order used by the main text. The final benchmark contains 129 tasks retained from 300 algorithm-selected candidates through automated screening and targeted human review. Component-validation samples and operational score summaries retain their measured denominators.

Appendix~\ref{app:data-construction} expands the data construction in \S\ref{sec:trace-retrieval}, covering reporting populations, taxonomy, selection, and distribution bias. Appendix~\ref{app:recovery} details the recovery and replay protocol in \S\ref{sec:recovery}, while Appendix~\ref{app:rubric-scoring} provides the rubric and scoring specifications underlying \S\ref{sec:rubric}.

Appendix~\ref{app:experiment-plan} supplies the extended validation and model results for \S\ref{sec:experiments}, and Appendix~\ref{app:feedback} describes the revision protocol and initial paired results in \S\ref{sec:rq3}. Appendix~\ref{app:feedback-case-study} follows one request through feedback and repair. Appendix~\ref{app:full-profile-tables} summarizes the profile taxonomy; complete specifications and pipeline inputs and outputs are available in the \href{https://anonymous.4open.science/r/TRACE2EVAL-ICLR}{anonymous repository}.

\subsection{Data Construction and Distribution}
\label{app:data-construction}
\label{app:trace-retrieval}

\subsubsection{Trace Selection and Benchmark Construction}
\label{app:cohorts}

The construction process in \S\ref{sec:trace-retrieval} turns online interactions into independently evaluable coding tasks. Claude Sonnet 4.6 performs trace labeling and automated admission judgments. Coding-request identification reduces approximately 20,000 collected sessions to 14,603 code-related traces, from which distribution-aligned greedy selection draws 300 candidates. A second greedy stage combines multi-round voting under the admission rubric with the same distributional objective to select the final 129 tasks; human reviewers resolve borderline admission judgments. Table~\ref{tab:cohorts} summarizes these stages and their construction decisions.

\begin{table}[!htbp]
\centering\small
\caption{Trace counts and selection criteria at each benchmark-construction stage.}
\label{tab:cohorts}
\begin{tabularx}{\linewidth}{l r X}
\toprule
Stage & Size & Construction step \\
\midrule
Collected sessions & $\approx20{,}000$ & Desensitized online coding-agent traffic. \\
Coding population $\sC$ & 14,603 & Identify coding requests and assign profile labels. \\
Selected candidates $\sS$ & 300 & Apply distribution-aligned greedy selection to the coding population. \\
Final benchmark $\sB$ & 129 & Combine rubric-based voting and distribution constraints in a second greedy stage; review borderline cases. \\
\bottomrule
\end{tabularx}
\end{table}

\paragraph{Identify coding requests and summarize user queries.}
We begin with 20,000 desensitized sessions collected from online coding-agent traffic between June and July 2026 by randomly sampling approximately equal numbers per day. We label these real user traces for coding relevance and summarize their user queries. Requests involving code development, modification, or analysis form the coding population $\sC$, containing 14,603 traces. Each query summary captures the user's objective and explicit requirements from the interaction, providing a concise description for subsequent profile labeling and candidate selection.

\paragraph{Characterize the coding population with user profiles.}
The traces in $\sC$ receive complementary labels describing the intended activity, technical context, application setting, and requester role. Task-type axes \texttt{x1}/\texttt{x2} and technical-domain axes \texttt{z1}/\texttt{z2} distinguish broad families from finer activities and specializations. Usage scenario \texttt{y} and user role \texttt{w} distinguish requests with similar technical content but different purposes or user groups. This labeling step supplies the population proportions used for selection, preserving distinctions that a single coding-related label would miss. Using the taxonomy summarized in Appendix~\ref{app:full-profile-tables}, we label and select online traces and SWE-bench instances, finding that SWE-bench does not match the online traffic distribution (Appendix~\ref{app:alignment-discussion}).

\paragraph{Select a representative candidate set.}
Candidate selection and admission screening aim to assemble tasks with sufficient difficulty to reveal performance differences across agents while remaining aligned with real user-profile distributions. Algorithm~\ref{alg:align-select} addresses the distributional objective by selecting $\sS\subseteq \sC$ with $|\sS|=300$. Category quotas reflect the population proportions, and greedy selection jointly fills deficits across profile axes. The second greedy stage applies the admission rubric while retaining the distributional objective, jointly considering task difficulty, discriminative value, and profile alignment. Appendix~\ref{app:selection-algorithm} details the selection procedure, and Appendix~\ref{app:alignment-discussion} examines the profile distributions of the 300 selected traces.

\paragraph{Define the admission rubric for candidate screening.}
We develop the human-defined admission rubric through experiments on a small calibration set. Table~\ref{tab:admission-rules} summarizes the criteria used to screen the 300 selected candidates $\sS$ into the final benchmark $\sB$. The criteria favor substantive implementation or code-reasoning demands expected to distinguish agent capabilities, while checking task clarity, recoverability, and distinctiveness. Together with distribution-aligned selection, this screening aims to provide informative comparisons across the range of requests represented in real coding traffic. The artifact-quality rubric in Appendix~\ref{app:rubric} is used later to grade the agent's delivered result.

\begin{table}[!htbp]
\centering\small
\caption{Admission rubric used to screen selected candidates through automated voting and targeted human review.}
\label{tab:admission-rules}
\setlength{\tabcolsep}{4pt}
\renewcommand{\arraystretch}{1.15}
\begin{tabularx}{\linewidth}{@{}>{\raggedright\arraybackslash}p{.20\linewidth} >{\raggedright\arraybackslash}X@{}}
\toprule
Criterion & Admission decision rules \\
\midrule
Task clarity & \textbf{Retain:} The objective, constraints, and task boundary can be recovered from the interaction.\par\smallskip
\textbf{Exclude:} Fragments or continuation requests whose essential context cannot be recovered. \\
\addlinespace[4pt]
Task difficulty & \textbf{Retain:} Substantive implementation or code reasoning is expected to reveal differences in agent performance.\par\smallskip
\textbf{Exclude:} Trivial edits, pure information lookup, or setup-only requests without substantive coding work. \\
\addlinespace[4pt]
Input recoverability & \textbf{Retain:} Essential files and data are observable or can be reconstructed from available evidence.\par\smallskip
\textbf{Exclude:} Missing private inputs or insufficient evidence prevent constructing the required starting context. \\
\addlinespace[4pt]
Environment feasibility & \textbf{Retain:} Required execution can be represented in the supported environment.\par\smallskip
\textbf{Exclude:} Essential authenticated services, remote systems, or internal platforms cannot be reproduced. \\
\addlinespace[4pt]
Task distinctiveness & \textbf{Retain:} The request contributes a distinct task objective or is the retained representative of a duplicate group.\par\smallskip
\textbf{Exclude:} Another retained request already covers an equivalent core objective and expected result. \\
\bottomrule
\end{tabularx}
\end{table}

\paragraph{Combine admission voting with distribution-constrained greedy selection.}
Across screening rounds, each candidate in $\sS$ receives five independent automated keep-or-remove judgments under the admission rubric. At least three agreeing votes determine the initial keep-or-remove decision, while cases near the voting threshold are referred for human reinspection. Reviewers examine these cases against the same criteria and resolve the admission decision, focusing human effort on ambiguous requests without requiring a full manual pass over all 300 candidates. Semantic deduplication retains representatives of equivalent task objectives. The second greedy stage combines these admission decisions with population-profile constraints to retain $\sB\subseteq \sS$ with $|\sB|=129$, preserving distribution alignment while satisfying the admission criteria.

\paragraph{Use trace labels to guide downstream recovery.}
Alongside user-profile categories, trace labeling provides initial judgments of whether a request is code-related and whether it requires workspace or environment recovery. These judgments draw on information in the original interaction and give downstream agents a preliminary reference for identifying the task's file and runtime requirements. The retained traces, query summaries, and labels are then passed to the recovery workflows in Appendix~\ref{app:recovery}. As illustrated in Figure~\ref{fig:recovery-workflow}, specialized agents use this information to recover the required files and reconstruct the execution environment, followed by trajectory replay to validate the recovered context.

\subsubsection{Profile Taxonomy and Reference Distributions}
\label{app:taxonomy-tables}

The \href{https://anonymous.4open.science/r/TRACE2EVAL-ICLR/profiles/README.md}{anonymous repository} provides the 171-category dictionary used in \S\ref{sec:trace-retrieval} and the full traffic and SWE-bench distribution tables. The traffic table concerns all 20,000 collected sessions, not only the 14,603 coding sessions. The SWE-bench comparison pools 1,531 instances from the three test splits listed in Table~\ref{tab:swe-comparison-sources}, each characterized under the same profile taxonomy. Category labels are translated for presentation without changing their identities.

The task-type hierarchy describes what users ask an agent to do. Axis \texttt{x1} groups broad activities such as task execution, programming, and information retrieval, while \texttt{x2} distinguishes operations such as feature implementation, bug fixing, code understanding, and automation orchestration. These predefined categories provide a consistent vocabulary for recurring activities across domains. Online requests, however, can combine several objectives or involve specialized workflows that do not map cleanly to a single activity category, even when their application setting or technical domain is identifiable. Allowing an unassigned task type accommodates this open-ended scope without forcing a category assignment. N/A records the absence of an assigned task-type label; the remaining profile axes can still characterize the request.

In the full traffic table, task-type labels are unassigned for 10,354 sessions (51.77\%) on each of \texttt{x1} and \texttt{x2}. We omit these N/A entries and normalize each task-type distribution over its 9,646 labeled sessions, so the reported proportions describe labeled tasks rather than all traffic. The other four axes retain N/A and use all 20,000 sessions as their denominator. Counts remain observed category frequencies. The coding-only comparisons in Figures~\ref{fig:profiles-task}, \ref{fig:profiles-context}, and~\ref{fig:profiles-domain} use the same label export restricted to code-related sessions; their task hierarchy likewise excludes unassigned labels and uses its own labeled-sample denominator.

We distinguish presence coverage from frequency alignment in Appendix~\ref{app:alignment-discussion}. Categories aggregated as Others in a compact figure remain individually represented in the repository tables.

\subsubsection{Distribution-Aligned Candidate Selection}
\label{app:selection-algorithm}

The selection step in Appendix~\ref{app:trace-retrieval} selects 300 candidates from the 14,603 code-related traces for subsequent screening under Table~\ref{tab:admission-rules}. Each coding trace carries labels for task type, usage scenario, technical domain, and user role, so selecting it changes several profile distributions at once. Algorithm~\ref{alg:align-select} coordinates these choices through category quotas and iterative deficit reduction. This is the first of two greedy stages; the second combines rubric-based admission with distribution constraints to select the final benchmark.

\paragraph{Translate profile proportions into target counts.}
Let $\sC$ denote the 14,603 code-related traces, $\sS\subseteq \sC$ the selected candidate set, and $n_{\mathrm{sel}}=300$ the selection budget. Each trace $\tau\in \sC$ has a label $\ell_a(\tau)$ on each profile axis $a\in\{\mathtt{x1},\mathtt{x2},\mathtt{y},\mathtt{z1},\mathtt{z2},\mathtt{w}\}$. For category $v$ on axis $a$, $p_{a,v}$ is its empirical proportion in $\sC$, and $c_{a,v}$ counts its occurrences in the current set $\sS$. The fractional allocation $q_{a,v}=n_{\mathrm{sel}}p_{a,v}$ gives the desired number of selected traces carrying that label. We obtain integer targets $t_{a,v}$ by taking the floor of each allocation and assigning the remaining seats to the largest fractional remainders. The targets on each axis sum to $n_{\mathrm{sel}}$. Because each trace contributes labels on several axes, these targets guide a joint approximation to the population distributions.

\paragraph{Select traces that fill several deficits.}
At each iteration, we score every unselected trace by the remaining target fractions of its labels,
\begin{equation}
\label{eq:selection-deficit}
\operatorname{Deficit}(\tau)=\sum_a
\frac{\bigl(t_{a,\ell_a(\tau)}-c_{a,\ell_a(\tau)}\bigr)^+}
{\max\!\bigl(t_{a,\ell_a(\tau)},1\bigr)},
\qquad \tau\in \sC\setminus \sS,
\end{equation}
where $(x)^+=\max(x,0)$. A category contributes its unfilled target fraction until its count reaches the target, after which its contribution is zero. Normalization makes relative shortfalls comparable across common and rare categories, and the denominator remains well defined when rounding gives a zero target. A trace can therefore address several shortfalls at once, such as an underrepresented task type and usage scenario. Algorithm~\ref{alg:align-select} selects the trace with the largest score, updates the counts of all its labels, and recomputes priorities for the next iteration. This procedure builds a set of 300 candidates while balancing the six profile axes without separately sampling every combination of labels.

\begin{algorithm}[H]
\caption{Distribution-aligned candidate selection.}
\label{alg:align-select}
\begin{algorithmic}[1]
\Require labeled coding population $\sC$ with labels $\ell_a(\tau)$; population proportions $p_{a,v}$; budget $n_{\mathrm{sel}}=300$
\Ensure candidate set $\sS$ and its per-axis category proportions
\For{each axis $a$}
\State $q_{a,v}\gets n_{\mathrm{sel}}p_{a,v}$ and $t_{a,v}\gets\floor{q_{a,v}}$ for each category $v$
\State Assign the remaining $n_{\mathrm{sel}}-\sum_vt_{a,v}$ seats by descending fractional remainder
\EndFor
\State $\sS\gets\emptyset$; $c_{a,v}\gets0$ for every axis $a$ and category $v$
\While{$|\sS|<n_{\mathrm{sel}}$ and $\sC\setminus \sS\ne\emptyset$}
\State Compute $\operatorname{Deficit}(\tau)$ for each $\tau\in \sC\setminus \sS$ using Equation~(\ref{eq:selection-deficit})
\State $\tau^\star\gets\argmax_{\tau\in \sC\setminus \sS}\operatorname{Deficit}(\tau)$, breaking ties by a fixed trace ordering
\State $\sS\gets \sS\cup\{\tau^\star\}$
\State $c_{a,\ell_a(\tau^\star)}\gets c_{a,\ell_a(\tau^\star)}+1$ for each axis $a$
\EndWhile
\State Return $\sS$ and proportions $c_{a,v}/|\sS|$ for every axis $a$ and category $v$
\end{algorithmic}
\end{algorithm}

\paragraph{Retain distribution constraints during admission.}
The algorithm returns the 300 selected candidates $\sS$, which then enter the admission workflow described in Appendix~\ref{app:trace-retrieval}. Using the rubric in Table~\ref{tab:admission-rules}, five independent automated judgments assess each candidate's suitability for informative agent comparisons, and human reviewers resolve borderline cases. A second greedy stage combines multi-round admission decisions with the same population-profile objective to select the 129-task benchmark $\sB\subseteq \sS$. Both stages therefore preserve distribution alignment while the second also enforces task difficulty and feasibility. To evaluate distribution alignment, we compare the candidates and SWE-bench with the same coding population under the same taxonomy; Appendix~\ref{app:alignment-discussion} presents the distributions in Figures~\ref{fig:profiles-task}, \ref{fig:profiles-context}, and~\ref{fig:profiles-domain} and their distances in Table~\ref{tab:historical-selection-alignment}.

\subsubsection{User-Profile Coverage and Distribution of the Candidate Set}
\label{app:alignment-discussion}

\begin{figure}[!htbp]\centering
\includegraphics[width=\linewidth]{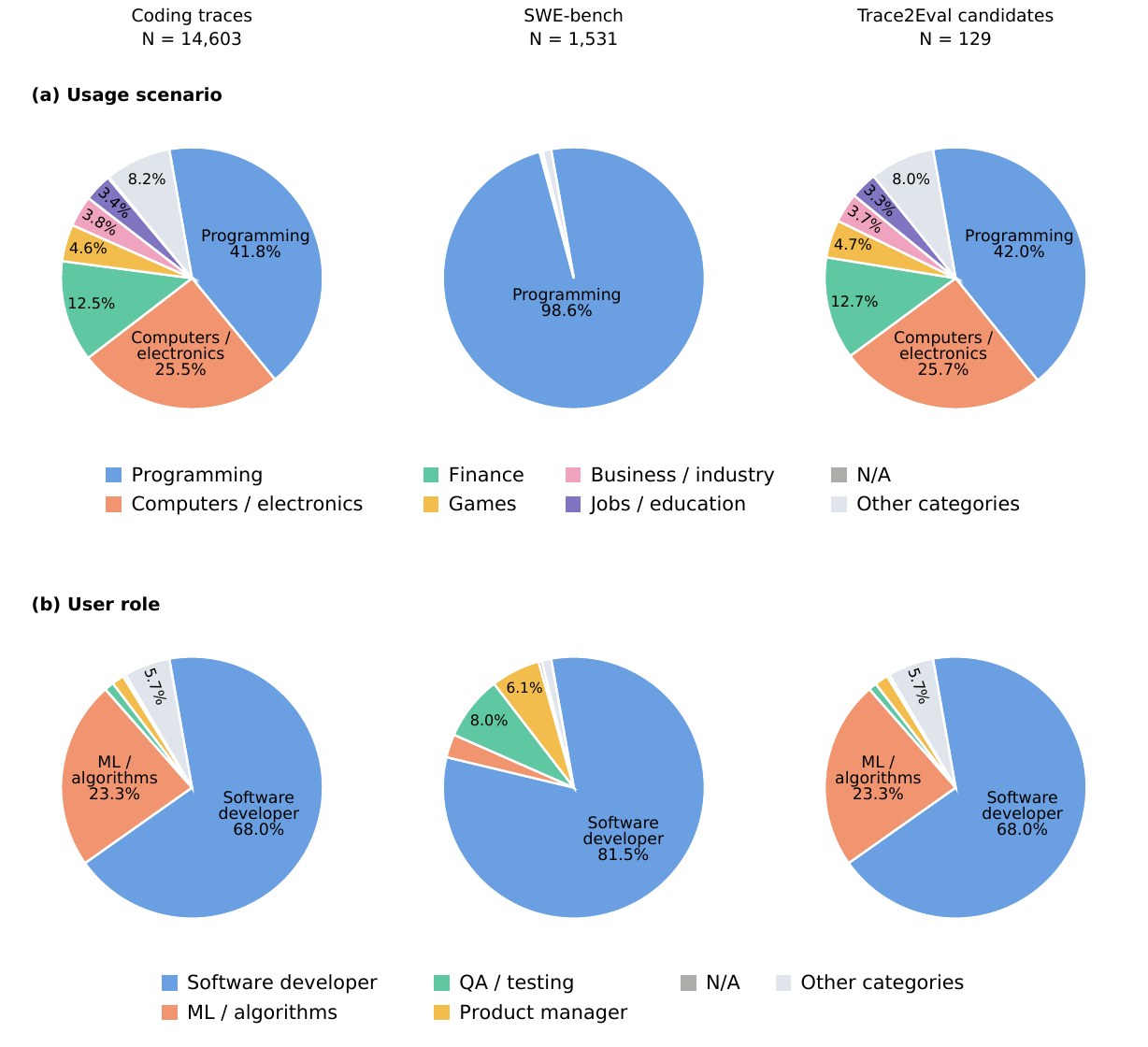}
\caption{Usage scenarios (a) and user roles (b) in coding traces, SWE-bench, and selected candidates. The same colors identify corresponding categories across the three populations.}
\label{fig:profiles-context}
\end{figure}

\begin{figure}[!htbp]\centering
\includegraphics[width=\linewidth]{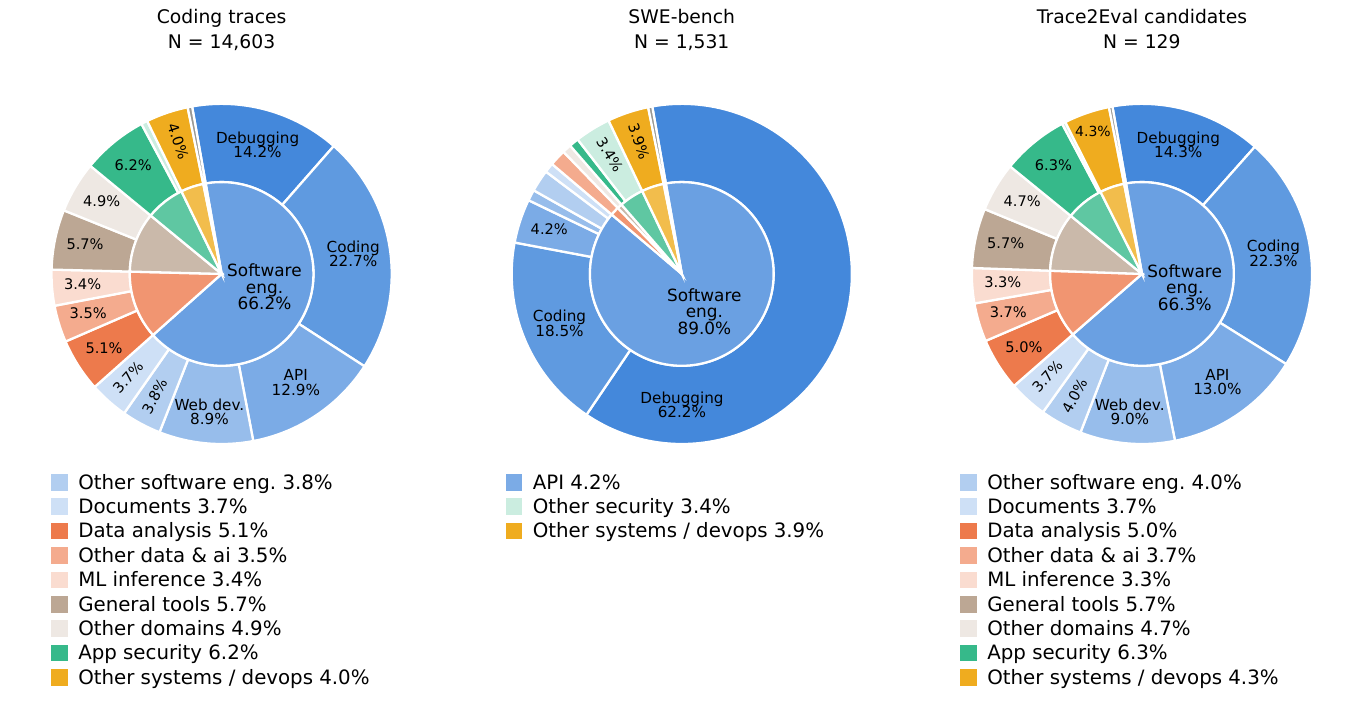}
\caption{Technical-domain profiles of coding traces, SWE-bench, and selected candidates. Inner disks show broad domains, and outer rings show activities within them.}
\label{fig:profiles-domain}
\end{figure}

We evaluate the distribution-matching step in Appendices~\ref{app:trace-retrieval} and \ref{app:selection-algorithm} using the full coding population $\sC$ of 14,603 traces. Applying Algorithm~\ref{alg:align-select} at a budget of 300 produces the candidate set $\sS$ for this analysis. We compare both $\sS$ and a pooled SWE-bench comparison set $\sW$ against the same population $\sC$ under the taxonomy in Appendix~\ref{app:taxonomy-tables}. Here, $\sW$ pools the test instances from SWE-bench Verified, Multilingual, and Pro. Together, they contribute 1,531 instances, as detailed in Table~\ref{tab:swe-comparison-sources}. Each SWE-bench instance contributes once to the pooled distribution, with its problem statement labeled using the same six profile axes as online requests. Figures~\ref{fig:profiles-task}, \ref{fig:profiles-context}, and~\ref{fig:profiles-domain} compare the resulting task, context, and technical-domain distributions. These visual comparisons characterize coverage of common requests and differences in their relative frequencies; the candidate set spans a broader range of observed activities than the SWE-bench comparison set.

\begin{table}[!htbp]
\centering\small
\caption{SWE-bench test sets pooled for the user-profile comparison.}
\label{tab:swe-comparison-sources}
\begin{tabular}{llr}
\toprule
Source benchmark & Split & Instances \\
\midrule
\href{https://huggingface.co/datasets/princeton-nlp/SWE-bench_Verified}{SWE-bench Verified} & Test & 500 \\
\href{https://huggingface.co/datasets/SWE-bench/SWE-bench_Multilingual}{SWE-bench Multilingual} & Test & 300 \\
\href{https://huggingface.co/datasets/ScaleAI/SWE-bench_Pro}{SWE-bench Pro} & Public test & 731 \\
\midrule
Combined comparison set $\sW$ & & 1,531 \\
\bottomrule
\end{tabular}
\end{table}

\paragraph{Measuring distribution alignment.}
We denote the individual benchmarks by $\sW_{\mathrm{Verified}}$, $\sW_{\mathrm{Multi}}$, and $\sW_{\mathrm{Pro}}$, and retain $\sW$ for their pooled set. Figures~\ref{fig:profiles-task}, \ref{fig:profiles-context}, and~\ref{fig:profiles-domain} show the pooled SWE-bench distribution as an overview, while Table~\ref{tab:historical-selection-alignment} additionally compares each benchmark separately. For any comparison set $\sT$, let $n^{\sT}_{a,v}$ count traces with category $v$ on axis $a$, and let $p^{\sT}_{a,v}=n^{\sT}_{a,v}/|\sT|$ be the corresponding proportion. With $p^{\sC}_{a,v}$ defined analogously for the coding population, total variation measures their distributional distance,
\begin{equation}
\label{eq:profile-total-variation}
\operatorname{TV}_a(\sC,\sT)=\frac12\sum_v\left|p^{\sT}_{a,v}-p^{\sC}_{a,v}\right|.
\end{equation}
The sum covers the same categories for both sets, assigning zero to categories absent from a set. The factor of one half avoids counting both the excess and the shortfall of the same probability mass. TV ranges from zero for identical distributions to one for disjoint distributions; a value of 0.02 means that 2\% of the probability mass would need to move between categories to match the reference. Table~\ref{tab:historical-selection-alignment} uses the full per-axis category counts and sample-set denominators, before any grouping used for visualization. This common reference makes TV a direct test of whether each set represents user activities at the frequencies observed in real traffic.

\paragraph{Covering task activities and user contexts.}
Figure~\ref{fig:profiles-task} shows why repository benchmarks offer a selective view of online demand. Among traces with assigned task types, programming accounts for 100.0\% of the SWE-bench comparison set, versus 53.1\% of coding traffic and 53.2\% of selected candidates. Bug fixing alone contributes 1,032 of 1,531 SWE-bench instances (67.4\%), compared with 1,046 of 7,590 labeled coding traces (13.8\%) and 22 of 158 labeled candidates (13.9\%). The selected set includes both programming and other common activities. Coverage also extends across user contexts. Programming scenarios account for 98.6\% of SWE-bench, but 41.8\% of coding traffic and 42.0\% of candidates. Likewise, software developers constitute 81.5\% of SWE-bench profiles, compared with 68.0\% in both the coding population and the selected set. SWE-bench therefore provides a narrower view of the observed mix of tasks, application contexts, and user roles. The candidate set covers a broader range of common requests.

\begin{figure}[!htbp]\centering
\includegraphics[width=\linewidth]{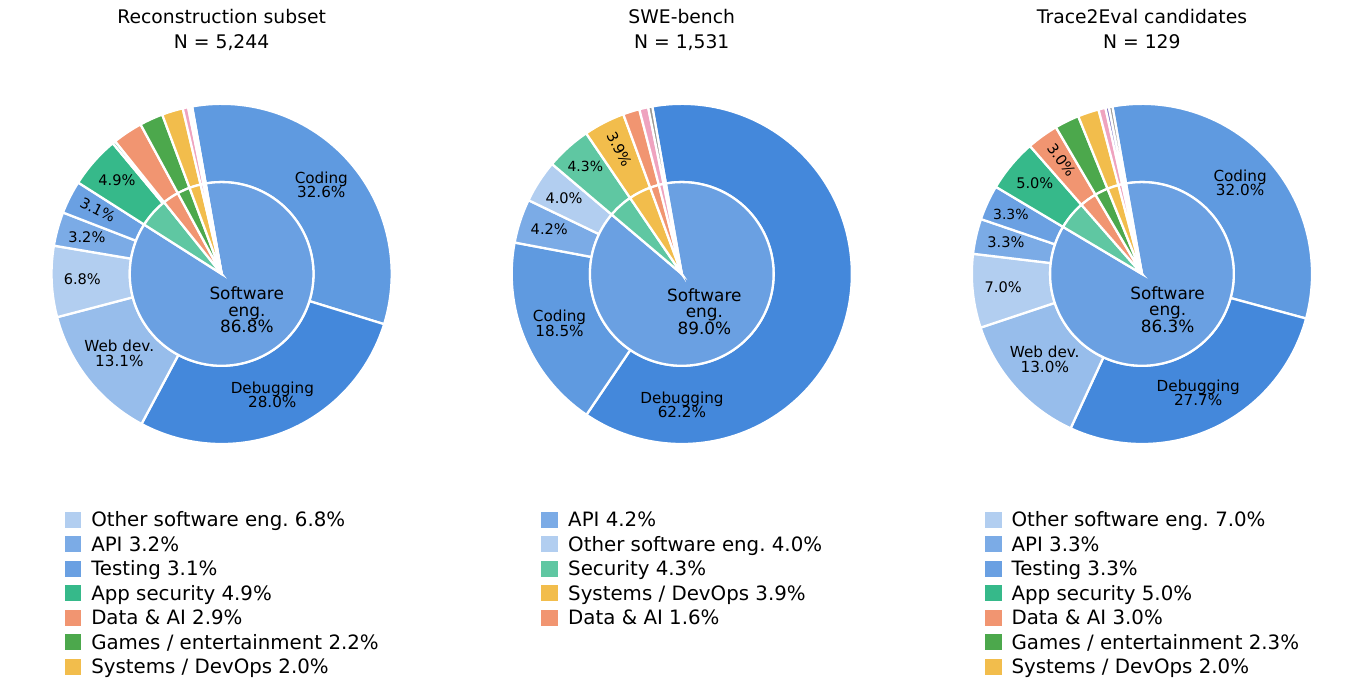}
\caption{Technical-domain profiles of the reconstruction subset, SWE-bench, and selected candidates. Inner disks show broad domains, and outer rings show activities within them.}
\label{fig:profile-alignment}
\end{figure}

Figure~\ref{fig:profile-alignment} summarizes the technical-domain hierarchy.


\begin{samepage}
\paragraph{Preserving broad domains and specific activities.}
Figure~\ref{fig:profiles-domain} shows that SWE-bench is misaligned with online demand at both domain and activity levels, overrepresenting software engineering and debugging while underrepresenting web development. Debugging accounts for 62.2\% of SWE-bench, compared with 14.2\% of coding traffic and 14.3\% of \method candidates. The candidates include these common domains and activities, extending coverage beyond the debugging-heavy profile of SWE-bench.
\par\end{samepage}

\paragraph{Alignment across all six profile axes.}
Table~\ref{tab:historical-selection-alignment} quantifies the contrast in Figures~\ref{fig:profiles-task}, \ref{fig:profiles-context}, and~\ref{fig:profiles-domain}: the 300-candidate set has lower measured distributional distance from the coding population than SWE-bench on all six axes. Candidate TV ranges from 0.0060 to 0.0232, compared with 0.2287 to 0.7877 for pooled SWE-bench. The candidates also have lower TV than every individual SWE-bench variant on every axis. These distances describe the algorithm-selected candidate set before admission screening.

\begin{table}[t]\centering\small
\setlength{\tabcolsep}{4pt}
\caption{Total variation (TV) between each comparison set and the coding population $\sC$ across six profile axes. Lower values indicate closer alignment. $\sW$ combines the three SWE-bench variants; $\sS$ denotes the selected \method candidates.}
\label{tab:historical-selection-alignment}
\begin{tabular}{lrrrrr}\toprule
 & \multicolumn{5}{c}{$\operatorname{TV}_a(\sC,\sT)\;\downarrow$} \\
\cmidrule(lr){2-6}
Profile axis & $\sW_{\mathrm{Verified}}$ & $\sW_{\mathrm{Multi}}$ & $\sW_{\mathrm{Pro}}$ & $\sW$ (total) & $\sS$ \\
\midrule
\texttt{x1} - Task type L1 & 0.7241 & 0.7241 & 0.7241 & 0.7241 & \textbf{0.0120} \\
\texttt{x2} - Task type L2 & 0.8453 & 0.8114 & 0.7887 & 0.7877 & \textbf{0.0191} \\
\texttt{y} - Usage scenario & 0.5795 & 0.5815 & 0.5664 & 0.5721 & \textbf{0.0220} \\
\texttt{z1} - Technical domain L1 & 0.2736 & 0.2846 & 0.2252 & 0.2287 & \textbf{0.0068} \\
\texttt{z2} - Technical domain L2 & 0.6932 & 0.7341 & 0.4551 & 0.5318 & \textbf{0.0232} \\
\texttt{w} - User role & 0.2224 & 0.2784 & 0.2733 & 0.2492 & \textbf{0.0060} \\
\bottomrule
\end{tabular}\end{table}

The observed user profiles guide coverage of common requests during candidate selection. The second greedy stage combines multi-round rubric voting with distribution constraints to produce the 129-task benchmark, jointly preserving profile alignment and evaluation value. The resulting benchmark is intended to cover diverse real-world coding needs; the artifact rubrics in Appendix~\ref{app:rubric} assess delivery quality within those task contexts.

\subsection{Workspace and Environment Recovery}
\label{app:recovery}

\begin{figure}[t]\centering
\includegraphics[width=\linewidth]{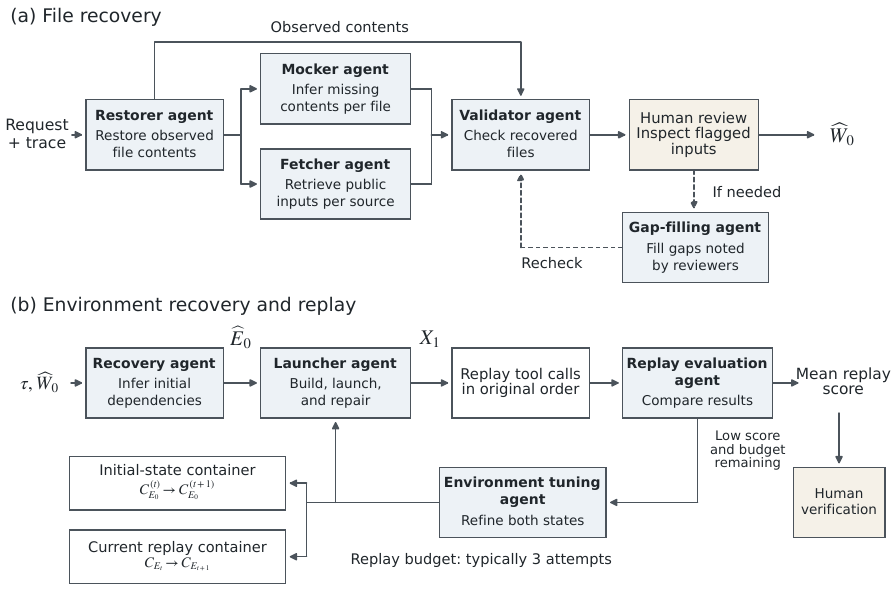}
\caption{Workspace and environment recovery. (a) Recovery agents restore observed file contents, infer missing contents, retrieve public inputs, and validate the assembled workspace, with human review guiding further corrections. (b) Tool calls are replayed in order, and their results guide updates to the initial and current environment states. Replay is limited to a preset budget, typically three attempts, before human verification. Here $\tau$ denotes the recorded trajectory and $X_1=(\widehat W_0,\widehat E_0)$ the recovered starting context.}
\label{fig:recovery-workflow}
\end{figure}

Recovery turns a retained user request into an executable task by reconstructing its initial files and dependencies where needed. We use Claude Opus 4.6 for the file and environment recovery workflows. An agent first assesses recovery needs from the previously summarized query and supporting evidence it locates in the trace, assigning each task to no recovery, file recovery only, or both file and environment recovery. When both are needed, files are recovered first so that environment inference uses the restored workspace. Following \S\ref{sec:recovery}, $q$ denotes the request and $\tau=(T_1,\ldots,T_T)$ its tool trajectory, with $T_t=(u_t,i_t,o_t)$ recording the tool, input, and observed output. The targets $W_0$ and $E_0$ are the workspace and environment before the recorded solution begins; hats denote recovered estimates. Figure~\ref{fig:recovery-workflow} shows how file recovery supplies $\widehat W_0$, environment inference supplies $\widehat E_0$, and replay checks their joint support for the recorded actions.

\subsubsection{File Recovery}
\label{app:file-recovery}

\paragraph{Restore observed inputs and identify missing contents.}
The restorer agent first extracts pre-existing files from tool reads, user-provided text, and embedded context. For files modified during the session, it preserves the observed earlier contents; files created or downloaded as part of the recorded solution are excluded from the initial workspace. A shared path mapping preserves directory relationships when moving inputs into the recovered workspace. This gives subsequent agents a common reference for connecting file evidence, imports, and tool arguments.

The resulting inventory distinguishes fully observed inputs, partially observed inputs, missing task-relevant contents, potential public resources, and task-irrelevant placeholders. It also identifies supporting files implied by imports or configuration references. Fully observed files require no generation. Missing contents are assigned either to contextual completion or to public retrieval, while unrelated files can remain empty placeholders. This division directs reconstruction effort toward inputs that affect the user's task.

\paragraph{Complete missing contents with a mocker agent.}
A mocker agent operates on one assigned file, using the user request $q$, its recorded tool trajectory $\tau$, and the restored files as context. For a partially observed file, it fills the missing portions while preserving the visible contents. For a missing task-relevant file, it reconstructs a compatible implementation or input from referenced interfaces, fields, imports, and usage. Public templates can inform standard configuration or library usage, whereas project-specific contents are inferred from the surrounding evidence. Other files remain read-only so that parallel completion does not overwrite the shared evidence.

More formally, let $f_{\mathrm{obs}}$ be the observed initial contents of a file at positions $\sO_f$, and let $\widehat f$ be its completed version. The preservation constraint is
\begin{equation}
\widehat f\big|_{\sO_f}=f_{\mathrm{obs}},
\label{eq:file-evidence-preservation}
\end{equation}
where restriction to $\sO_f$ selects the observed portions. Complete observations fix the entire file; partial observations constrain only the visible portions. The remaining contents must fit the request and neighboring interfaces, and are recorded as inferred rather than directly recovered.

\paragraph{Retrieve public datasets and resources.}
The fetcher agent handles inputs such as public datasets, model weights, and resource bundles whose authentic contents can be obtained externally. Unlike the mocker agent, which is dispatched per file, the fetcher agent works on files grouped by source so that related inputs share the same dataset or resource version. It first verifies the source using names, versions, and usage in the trace, then retrieves the required files and checks their placement, format, and whether the downloaded files are nonempty.

Source verification also determines whether retrieval is appropriate. A user-modified file or an unsupported public-source attribution is reassigned to contextual completion when its contents can be inferred. Failed downloads and unavailable task-critical private inputs remain explicit gaps. Thus, public retrieval preserves authentic data where available, while outputs from the mocker agent supply inferred context without being mistaken for original dataset contents. Independent file completions and source-group retrievals can proceed in parallel before their outputs are merged.

\paragraph{Validate the assembled workspace and resolve review findings.}
The validator agent checks the recovered inventory against the files on disk and inspects key files for compatible imports, exports, configuration references, and meaningful contents. Evidence records connect restored portions to their trace observations and distinguish inferred contents from retrieved resources. These records support targeted inspection of uncertain inputs. When human review identifies a missing input, a separate gap-filling agent uses the reviewer notes and original trace to supplement the existing workspace, followed by checks of the requested additions and syntax. The resulting $\widehat W_0$ is the fixed starting snapshot for environment recovery and replay. The acceptance study checks recovered files against the request and trace evidence, recording missing inputs and unintended solution artifacts; Table~\ref{tab:workspace-calibration} in Appendix~\ref{app:recovery-calibration} gives the results.

\subsubsection{Environment Recovery and Replay}
\label{app:environment-recovery}

\paragraph{Infer the initial environment through temporal evidence.}
The recovery agent combines $\widehat W_0$ with installation commands, runtime indicators, errors, service usage, and configuration changes in $\tau$. A feasibility check first identifies which platform dependencies can be represented in the execution environment. The update in Eq.~\ref{eq:env-step} maintains two container states throughout replay. $C_{E_t}$ is the current environment before call $t$, and $C_{E_0}^{(t)}$ is the initial-environment estimate after the preceding calls in that round. The current container begins each round from the latest initial-state estimate. Processing $T_t$ advances the current environment and provides evidence for correcting both containers: missing initial dependencies require changes to the initial-state container as well as the current replay state. The index $t$ follows tool calls within a round; subsequent rounds repeat this process with the revised initial environment.

Successful use of a dependency before any installation supports its inclusion in $\widehat E_0$. An explicit later installation instead belongs to the inferred change $\widehat{\Delta E}_t$ at that step. For example, a package first installed midway through the trace should become available at that point during replay, rather than being preinstalled in the initial image. The initial-state container $C_{E_0}^{(T+1)}$ at the end of a round realizes $\widehat E_0$, while inferred changes between successive current states define $\widehat{\Delta E}_t$. Thus, $\widehat E=(\widehat E_0,\widehat{\Delta E}_1,\ldots,\widehat{\Delta E}_T)$ describes the recovered starting environment and its subsequent evolution. Initial requirements are mapped to a supported Linux configuration and packaged into a container. A launcher agent builds and cold-starts it, repairs configuration issues revealed by build or startup failures, and checks that it is ready for replay.

\paragraph{Replay the trajectory from the recovered context.}
Since a complete ground-truth workspace and environment are unavailable, we assess recovery through trajectory equivalence: agreement between replayed tool-call results and their recorded observations. First, we extract an ordered list of tool calls from $\tau$, retaining each tool $u_t$, input $i_t$, and original observation $o_t$. Second, we cold-start the recovered environment $\widehat E_0$ with a working copy of $\widehat W_0$ and replay the calls in their original order. Third, we compare each eligible replay observation $a_t$ with its recorded counterpart $o_t$ and aggregate the per-call agreement scores. These comparisons test recovery fidelity and locate discrepancies for subsequent refinement.

Within a replay round, let $X_t$ denote the joint workspace and current environment state immediately before call $t$, with $X_1=(\widehat W_0,\widehat E_0)$. Its environment component is $C_{E_t}$; the initial-state container $C_{E_0}^{(t)}$ is maintained separately. For an executed action,
\begin{equation}
(X_{t+1},a_t)=\operatorname{Exec}\bigl(X_t,u_t,\rho(i_t)\bigr),
\label{eq:replay-transition}
\end{equation}
where $\rho$ applies the shared path mapping to tool inputs and $a_t$ is the actual observation. Executed writes, installations, and service operations update the replay state in trajectory order. The changes $\widehat{\Delta E}_t$ explain these transitions and are not additional actions applied a second time. Replay modifies a working copy of the workspace, preserving the recovered file snapshot while allowing both environment containers to be refined.

\paragraph{Compute the replay score for recovery acceptance.}
For each replayed tool call, a replay evaluation agent compares the result $a_t$ produced in the recovered workspace and environment with the original result $o_t$, using the tool $u_t$ and its input $i_t$ as context. It assigns an agreement score $m_t\in[0,1]$: identical results receive 1, while increasingly different results receive lower scores, approaching 0 for substantially different outcomes. These judgments form the trajectory-equivalence rubric; their arithmetic mean is the \emph{replay score} in Eq.~\ref{eq:equiv-main}. A score of 1 after replaying and scoring the complete tool-call sequence indicates an equivalent working context for the recorded trajectory. We use this score to assess overall recovery quality for acceptance, while low-scoring calls locate discrepancies for refinement.

\paragraph{Use replay discrepancies to refine the recovered context.}
The automated refinement loop has a preset replay budget, typically three attempts. In each round, low-scoring tool calls identify file or environment discrepancies for targeted repair. Recovery agents correct the affected workspace files, while the environment tuning agent updates both the initial-state and current-state containers. When discrepancies remain and the budget permits, the launcher agent rebuilds and cold-starts the revised initial environment with the updated recovered workspace for another round. Each round compares tool-call results against the same recorded observations to verify these corrections. The resulting context and replay evidence then undergo human verification for recovery acceptance, as shown in Figure~\ref{fig:overview}. The recovery studies assess acceptance (Table~\ref{tab:rq1-combined}), compare single-shot recovery, launch repair, and full recovery (Table~\ref{tab:rq1-env-ablation}), and measure replay scores before and after refinement on the same cases; Appendix~\ref{app:environment-calibration} analyzes the outcomes.

\subsection{Rubric and Scoring Protocol}
\label{app:rubric-scoring}

\subsubsection{Rubric Design and Human Annotation}
\label{app:rubric}

Open-ended coding requests admit multiple valid implementations, so artifact quality must be judged against the user's needs rather than a unique reference solution. The rubric in \S\ref{sec:rubric} separates accuracy, code quality, and user experience into eleven dimensions. Its construction follows two stages on calibration artifacts. We first refine the human criteria to improve agreement between annotators and establish a stable reference, then refine the agent's scoring points to improve agreement with that fixed human reference. Ground truth (GT) denotes the final human reference from a third annotation round that integrates the first two annotators' scores and rationales with direct artifact inspection.

\paragraph{Ground the dimensions in human judgments.}
\label{app:rubric-basis}
The dimensions are informed by a qualitative analysis of human annotations and the written reasons for preferring one coding output over another. Table~\ref{tab:rubric-basis} summarizes the concerns expressed in those rationales. A concern's frequency is the proportion of reviewed annotation results whose rationale mentions that concern, using the same collection of annotations as the denominator for every row. It measures how often a consideration enters a human judgment.

A rationale can cite several concerns, such as incomplete functionality together with a high repair cost, so the categories overlap and their proportions need not sum to 100\%. Related concerns can also motivate the same rubric dimension. Functional completeness appears in 58.5\% of the reviewed rationales and executability in 33.8\%, supporting separate checks of whether the requested functionality is implemented and whether the artifact runs. Repair cost appears less frequently, but captures a different practical concern about how much work remains for the user. These frequencies inform the choice and scope of dimensions; they are not used as scoring weights.

\begin{table}[t]
\centering\small
\setlength{\tabcolsep}{5pt}
\renewcommand{\arraystretch}{1.12}
\caption{User-quality concerns identified in human annotation rationales and their corresponding rubric dimensions. Frequencies indicate how often each concern is mentioned; a rationale may include several concerns.}
\label{tab:rubric-basis}
\begin{tabularx}{\linewidth}{@{}>{\raggedright\arraybackslash}X >{\raggedright\arraybackslash}X r@{}}
\toprule
Human concern & Rubric dimension & Frequency (\%) \\
\midrule
Requested functionality & Requirement implementation & 58.5 \\
Executability and runtime setup & Code execution & 33.8 \\
Trustworthy output data & Data authenticity & 10.8 \\
Robustness and error handling & Robustness & 26.2 \\
Architecture and technology fit & Architecture and technical choice & 16.9 \\
Clarity of the solution and its logic & Readability; requirement implementation & 12.3 \\
Runtime and memory efficiency & Time/space complexity & 6.2 \\
Effort and risk of repair & Repair cost & 4.6 \\
Layout and interaction usability & Interface and interaction & 13.8 \\
Useful additional functionality & Feature richness & 10.8 \\
Organization of displayed information & Information presentation & 6.2 \\
Implementation-constraint compliance & Requirement implementation & 1.5 \\
Redundant or missing explanations & Readability; information presentation & 3.1 \\
\bottomrule
\end{tabularx}
\end{table}

Starting from these concerns, we collect expert judgments of how strongly each quality dimension affects user experience in real development. Experts assess the consequences of defects for task completion, usability, and subsequent maintenance or repair. These judgments establish group weights of 50\% for accuracy, 35\% for code quality, and 15\% for user experience, as well as the within-group priorities in Table~\ref{tab:rubric-dims}. Subsequent reassessment confirms the final weighting scheme used in the reported model scores. These weights preserve expert priorities in artifact-quality aggregation and human-reference agreement summaries and remain fixed during agent calibration.

\paragraph{Fix expert weights and evidence requirements.}
\label{app:rubric-weights}
Let $c$ index the three quality groups, $\lambda_c$ its group weight, and $\sD_c$ its dimensions. Each dimension $d$ has an expert-assigned relative priority $r_d$ within its group. We convert these priorities into the final dimension weights by normalization:
\begin{equation}
w_d=\lambda_c\frac{r_d}{\sum_{d'\in \sD_c}r_{d'}},
\qquad d\in \sD_c,\qquad \sum_d w_d=1.
\label{eq:rubric-weights}
\end{equation}
The relative priorities in Table~\ref{tab:rubric-dims} are normalized separately within each group. Requirement implementation receives $0.50\times45/100=0.225$ and code execution receives $0.50\times40/100=0.20$, preserving the expert priorities discussed in RQ2.1. These final weights govern the reassessed model Total in Table~\ref{tab:rq3-panel}, connecting expert judgments of user needs to the reported aggregate scores. Appendix~\ref{app:scoring} specifies their application to each artifact and the subsequent model-level average.

Evidence requirements determine how each criterion is assessed. Seven dimensions can be judged through source inspection. Code execution, data authenticity, interface and interaction, and information presentation additionally require runtime or rendered evidence. Relevant observations may inform more than one dimension, with dimension-specific overview instructions in Table~\ref{tab:rubric-anchors}.

\begin{table}[t]\centering\small
\setlength{\tabcolsep}{5pt}
\renewcommand{\arraystretch}{1.10}
\caption{Evidence requirements and final expert-informed weights for artifact scores and human-reference agreement summaries. Within each group, priorities $r_d$ are normalized and multiplied by the group weight $\lambda_c$ to obtain $w_d$.}
\label{tab:rubric-dims}
\begin{tabularx}{\linewidth}{@{}>{\raggedright\arraybackslash}X l r r@{}}
\toprule
Dimension & Evidence & Priority $r_d$ & Weight $w_d$ (\%) \\
\midrule
\multicolumn{4}{@{}l}{\textbf{Accuracy}\quad$\lambda_c=50\%$}\\
Requirement implementation & Static & 45 & 22.5 \\
Code execution & Dynamic & 40 & 20.0 \\
Data authenticity & Dynamic & 15 & 7.5 \\
\addlinespace[3pt]
\multicolumn{4}{@{}l}{\textbf{Code quality}\quad$\lambda_c=35\%$}\\
Readability & Static & 15 & 7.5 \\
Robustness & Static & 20 & 10.0 \\
Time/space complexity & Static & 10 & 5.0 \\
Repair cost & Static & 10 & 5.0 \\
Architecture and technical choice & Static & 15 & 7.5 \\
\addlinespace[3pt]
\multicolumn{4}{@{}l}{\textbf{User experience}\quad$\lambda_c=15\%$}\\
Interface and interaction & Dynamic & 15 & 7.5 \\
Information presentation & Dynamic & 10 & 5.0 \\
Feature richness & Static & 5 & 2.5 \\
\bottomrule
\end{tabularx}\end{table}

\paragraph{Derive agent scoring points from the human rubric.}
We first iterate the human score anchors until annotators achieve high agreement, then fix the rubric dimensions, weights, and human reference judgments. The judge agent initially receives this human rubric directly, but its scores do not align sufficiently with the human scores. We therefore iteratively revise the agent's scoring points over multiple rounds, evaluating each revision against the same human reference while keeping the dimensions and weights unchanged. The final agent anchors result from this calibration process.

Table~\ref{tab:rubric-anchors} summarizes the overview instructions; the \href{https://anonymous.4open.science/r/TRACE2EVAL-ICLR/rubric/score_anchors.md}{complete human and agent score anchors} are provided in the anonymous repository. Both retain the same three-level quality scale. The additional detail in the agent anchors reflects clarifications introduced to address observed disagreements, including what evidence to inspect, how to distinguish severity levels, and where dimension boundaries lie. Thus, the agent scoring points are calibrated instructions for applying the established human standard.

Human annotators can use professional experience and task context to interpret broad anchors, including whether a defect is consequential or a design fits the project. The agent needs more detailed guidance on these judgments, with explicit evidence requirements, severity thresholds, and boundary cases. These instructions help it apply the same quality standard without relying on superficial cues or counting one defect across multiple dimensions. To assess the calibrated rubric, we compare the two human annotators, the agent, and GT on the same artifacts; Table~\ref{tab:rq2-kappa} in Appendix~\ref{app:paired-judge-agreement} presents these agreement comparisons.

\begin{table}[t]
\centering\small
\setlength{\tabcolsep}{5pt}
\renewcommand{\arraystretch}{1.12}
\caption{Overview instructions for the eleven rubric dimensions. The complete human and agent score anchors are available in the anonymous repository.}
\label{tab:rubric-anchors}
\begin{tabularx}{\linewidth}{@{}>{\raggedright\arraybackslash}p{.25\linewidth} >{\raggedright\arraybackslash}X@{}}
\toprule
Dimension & Overview instruction \\
\midrule
\textbf{Requirement implementation} & Map every core requirement to its implementation location or identify its absence. \\
\textbf{Code execution} & Use actual startup, interaction, service-response, and output evidence. \\
\textbf{Data authenticity} & Trace material outputs to their input, parser, computation, or service, and compare with observed results where available. \\
\textbf{Readability} & Cite naming, control flow, decomposition, and specific code passages from the perspective of a developer taking over the artifact. \\
\textbf{Robustness} & Inspect error branches, input validation, resource lifetimes, and recovery behavior on core paths. Identify what happens after an exception is caught, including silent suppression or useful handling. \\
\textbf{Time/space complexity} & Inspect the main algorithm and high-frequency paths, explaining how time or memory grows with the workload. \\
\textbf{Repair cost} & First identify concrete remaining defects, then assess the affected code, dependencies, and scope of repair. \\
\textbf{Architecture and technical choice} & Describe the actual modules, dependencies, data flow, and technology choices. \\
\textbf{Interface and interaction} & Inspect actual rendered screenshots together with interaction evidence. Examine layout, entry points, action feedback, and completion of core user operations. \\
\textbf{Information presentation} & Inspect rendered information hierarchy, grouping, text and chart legibility, and adaptation to the display. \\
\textbf{Feature richness} & Separate explicitly requested capabilities from useful additions, such as filtering, search, statistics, or import and export. List each addition and locate its implementation, distinguishing working behavior from decorative controls and placeholders. \\
\bottomrule
\end{tabularx}
\end{table}

\paragraph{Apply evidence requirements and dimension boundaries consistently.}
Each grade must be supported by source, runtime, or visual evidence under the dimension-specific instructions in Table~\ref{tab:rubric-anchors}. The judge agent records missing evidence, distinguishes evaluation failures from artifact defects, and respects dimension boundaries to avoid penalizing the same defect repeatedly.

\paragraph{First stabilize human judgments.}
\label{app:annotation}
All annotators hold a bachelor's degree in a computer-related discipline and have at least two years of software development experience. The first two human annotation rounds are carried out independently by Human A and Human B on the same calibration artifacts. Each annotator applies the human rubric and records a score and supporting rationale for every artifact--dimension pair. We compare their scores and rationales to identify ambiguities in the anchors, revise the human rubric, and repeat independent assessment until high agreement is reached.

\paragraph{Establish GT through a third human annotation round.}
With the human rubric fixed, a third annotator reviews the first two annotators' scores and scoring rationales alongside the artifact itself. This review integrates the earlier assessments with direct inspection to assign the final reference score $g_{i,d}$ for artifact $i$ on dimension $d$. These third-round human scores constitute GT. The original independent scores and rationales remain available as Human A and Human B, supporting both direct inter-annotator comparisons and comparisons with the final human reference.

\paragraph{Then refine agent scoring points against the human reference.}
Starting from the human rubric, we repeatedly compare the judge agent's grades and rationales with GT and refine its evidence requirements, severity thresholds, and dimension boundaries. The dimensions, weights, and human reference remain fixed throughout; only the agent's scoring points change. Table~\ref{tab:rubric-anchors} summarizes the resulting inspection instructions. Appendix~\ref{app:scoring} defines the agreement measures, while Table~\ref{tab:rq2-kappa} and Figure~\ref{fig:judge-calibration} report the human--human, human--agent, and GT-referenced comparisons.

\subsubsection{Evidence Collection and Score Aggregation}
\label{app:scoring}\label{app:panel-scoring}

\begin{figure}[t]\centering
\includegraphics[width=\linewidth]{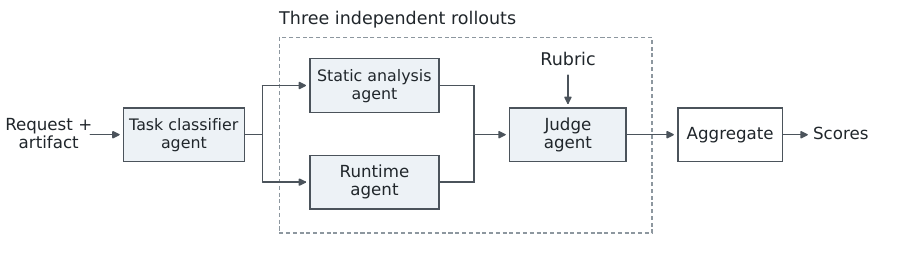}
\caption{Artifact scoring workflow. A task classifier agent selects the evaluation setting. In each of three independent rollouts, static analysis and runtime agents collect evidence, and a judge agent assigns rubric-based grades. These grades are aggregated into artifact scores and measures of agreement across rollouts.}
\label{fig:judge-workflow}
\end{figure}

The scoring workflow applies the calibrated agent criteria described in Appendix~\ref{app:rubric} to a candidate artifact. Figure~\ref{fig:judge-workflow} separates evidence collection, rubric-based judgment, and numerical aggregation. Static analysis and runtime agents collect evidence, which the judge agent uses to assign dimension-level grades under the fixed rubric. GLM-5.3-Flash serves as the judge agent for the model-panel artifacts. Deterministic aggregation converts repeated grades into an artifact score and diagnostics of agreement.

\paragraph{Collect evidence for the rubric dimensions.}
A task classifier agent uses the request and artifact structure to select the task-specific rubric and execution setting. A browser, application, service, script, or desktop workflow supplies observations appropriate to the task. A static analysis agent identifies implemented functions, missing components, and code relevant to the quality criteria. A runtime agent exercises the artifact and records outputs, failures, interaction states, and screenshots. It can configure the execution environment but leaves the candidate's source unchanged.

Source evidence supports the seven static dimensions in Table~\ref{tab:rubric-dims}; execution and rendered evidence additionally support code execution, data authenticity, interface and interaction, and information presentation. The two evidence sources are interpreted together. For example, finding an event handler establishes that an interaction is implemented, while exercising the control establishes whether it works for the user. The judge agent also distinguishes an observed artifact defect from a failure to obtain the required evidence, following the calibrated score anchors described in Appendix~\ref{app:rubric}.

\paragraph{Repeat observation and judgment independently.}
Let $i$ index an evaluated artifact and $d$ a rubric dimension. For each artifact, we use $K=3$ independent scoring rollouts, indexed by $k\in\{1,\ldots,K\}$. Each rollout collects its own source and runtime evidence without access to the observations or grades of the other rollouts. Thus, repetition captures variation in both the evidence found and its interpretation.

Following \S\ref{sec:rubric}, $e^{(k)}_{i,d}$ denotes the evidence relevant to dimension $d$ in rollout $k$, and $R_d$ is its fixed agent-facing specification, including the score anchors and evidence requirements. The judge agent $J$ assigns
\begin{equation}
\hat h^{(k)}_{i,d}=J\bigl(R_d,e^{(k)}_{i,d}\bigr)\in\{0,1,2\},
\label{eq:judge-step-app}
\end{equation}
where a larger grade indicates stronger delivery. Each grade is accompanied by a rationale connecting specific observations to the applicable anchor. $R_d$ is held constant across scoring rollouts; the rubric refinement described in Appendix~\ref{app:rubric} occurs during calibration, not while judging a benchmark artifact.

\paragraph{Separate the mean score, comparison label, and self-consistency.}
Let $\sK_{i,d}\subseteq\{1,\ldots,K\}$ contain the indices of rollouts that return a valid grade for artifact $i$ on dimension $d$, and let $K_{i,d}=|\sK_{i,d}|$ be their number. When $K_{i,d}>0$, aggregation gives
\begin{align}
\mu_{i,d}&=\frac{1}{K_{i,d}}\sum_{k\in\sK_{i,d}}\hat h^{(k)}_{i,d},
\label{eq:judge-mean-app}\\
a_{i,d}&=\operatorname{round}(\mu_{i,d}),
\label{eq:judge-label-app}\\
\mathrm{SC}_{i,d}&=\frac{1}{K_{i,d}}
\max_{v\in\{0,1,2\}}\sum_{k\in\sK_{i,d}}
\1\{\hat h^{(k)}_{i,d}=v\}.
\label{eq:judge-consistency-app}
\end{align}
Here $\mu_{i,d}\in[0,2]$ is the mean used to compute artifact quality, $a_{i,d}\in\{0,1,2\}$ is the rounded label used for comparison with human judgments, and $\mathrm{SC}_{i,d}$ is the fraction of available rollouts assigning the most frequent grade. The indicator $\1\{\cdot\}$ equals one when its condition holds and zero otherwise. Rounding uses the nearest integer, with ties to even.

For example, grades $(2,2,1)$ give a mean of $5/3$, a comparison label of $2$, and self-consistency of $2/3$. The mean preserves variation between grades, whereas self-consistency describes their reproducibility. A consistently repeated grade can still disagree with humans, so self-consistency does not replace human-reference calibration. If no rollout supplies a valid grade for a dimension, that dimension remains unscored.

\paragraph{Aggregate dimensions into artifact and model scores.}
Model comparison uses the final expert-informed weights $w_d$ in Table~\ref{tab:rubric-dims}, as applied in the reassessment underlying the reported Total scores. Let $\sD_i$ be the nonempty set of applicable dimensions with valid mean scores for artifact $i$. Its normalized score is
\begin{equation}
s_i=\frac{\sum_{d\in\sD_i}w_d\,\mu_{i,d}/2}{\sum_{d\in\sD_i}w_d}\in[0,1].
\label{eq:artifact-score-app}
\end{equation}
Dividing by two places each dimension on the unit interval; the denominator renormalizes the final weights over the applicable, scored dimensions while preserving their relative priorities. When all eleven dimensions apply and are scored, the denominator is one and accuracy, code quality, and user experience retain their group weights of 50\%, 35\%, and 15\%. Group summaries use the same weighted calculation restricted to the corresponding dimensions. Reported artifact and group scores multiply the normalized values by 100.

For model $m$, let $\sA_m$ contain the indices of its scored artifacts, with one artifact per evaluated task, and let $n_m=|\sA_m|$. For $n_m>0$, its mean normalized score and reported score are
\begin{equation}\label{eq:panel-mixture}
\bar s_m=\frac{1}{n_m}\sum_{i\in\sA_m}s_i,
\qquad \operatorname{Score}(m)=100\bar s_m.
\end{equation}
Each scored task therefore contributes equally to the model average. Appendix~\ref{app:model-panel-configuration} describes how model-panel results are consolidated across compute nodes, and Table~\ref{tab:panel-dimensions} reports dimension-level scores under this aggregation rule. Appendix~\ref{app:cost-effectiveness} compares quality with execution cost on common task sets. Revision comparisons match tasks and starting artifacts across rounds as described in Appendix~\ref{app:feedback}.

\paragraph{Compare the automated label with the human reference.}
The reference $g_{i,d}$ is the third-round human annotation defined in Appendix~\ref{app:annotation}. For a dimension $d$, let $\sP_d$ contain the indices of artifacts with both an automated comparison label $a_{i,d}$ and a human reference $g_{i,d}$, and let $N_d=|\sP_d|$. For $N_d>0$, exact agreement is
\begin{equation}
\mathrm{Exact}_d=\frac{1}{N_d}\sum_{i\in\sP_d}
\1\{a_{i,d}=g_{i,d}\}.
\label{eq:judge-exact-app}
\end{equation}
Within-one agreement replaces the equality condition with $|a_{i,d}-g_{i,d}|\le1$. Signed bias is the mean of $a_{i,d}-g_{i,d}$ over the same paired artifacts, with a positive value indicating more generous automated grades. Usability agreement compares $\1\{a_{i,d}>0\}$ with $\1\{g_{i,d}>0\}$ to measure agreement on the distinction between inadequate and at least partially usable delivery.

Cohen's $\kappa_d=(p_{\mathrm{o},d}-p_{\mathrm{e},d})/(1-p_{\mathrm{e},d})$ additionally adjusts for chance agreement. Here $p_{\mathrm{o},d}=\mathrm{Exact}_d$, and $p_{\mathrm{e},d}$ is the sum, over score categories, of the product of the agent and human-reference category proportions on $\sP_d$. It is undefined when $p_{\mathrm{e},d}=1$. This adjustment helps distinguish agreement produced by matching individual judgments from agreement driven by assigning one common grade to most artifacts.

Human--GT comparisons apply the same metrics separately to each annotator's original grades, then average the two annotator-level results. Agent--GT uses $a_{i,d}$ against the same reference. These comparisons implement the human-alignment objective in Appendix~\ref{app:rubric}, while self-consistency remains a diagnostic of repeated automated scoring. Weighted agreement summaries use the same final expert-informed weights as artifact scoring (Appendix~\ref{app:rubric-weights}), applying them to dimension-level agreement statistics rather than artifact grades. Table~\ref{tab:rq2-kappa} reports the resulting chance-corrected comparisons.

\subsection{Supplementary Validation and Agent Results}
\label{app:experiment-plan}

This section supplements the main experiments with component-level validation and extended analyses. Appendix~\ref{app:recovery} defines how recovery and trajectory-equivalence checks operate. Here we explain how the recovered files and environments are assessed, which cases enter each measurement, and what the observed results establish. The workspace recovery study checks recovered inputs directly, while the environment studies examine deployment feasibility, replay outcomes, and the contribution of repair. Subsequent analyses assess judge agent agreement, explain how recovered tasks support candidate-agent execution and artifact scoring, and describe model-result aggregation.

\subsubsection{Workspace recovery calibration}
\label{app:recovery-calibration}
\paragraph{Identify cases that need file recovery.}
We use a calibration set of 100 cases to assess recovery quality. Human annotators first label whether each task requires pre-existing local files, and the agent independently makes the same decision from the request and trace. Their decisions agree on 96 of the 100 cases; 57 cases require file recovery.

\paragraph{Reconstruct the files and assess their quality.}
The agent reconstructs the workspaces for these 57 cases, after which human annotators assess whether the required inputs and observed contents are restored faithfully. They check completeness, unsupported file generation, and the inclusion of files created during the original solution. Human reviewers accept 82.6\% of the assessable reconstructions as successful. Table~\ref{tab:workspace-calibration} reports the assessed case counts, acceptance rates, and file-quality findings; Table~\ref{tab:rq1-combined} summarizes workspace and environment recovery results.

\begin{table}[t]
\centering\small
\caption{Human assessment of workspace recovery. Agreement on recovery need compares agent and human decisions; reconstruction success measures acceptance of the recovered files. Each rate uses the cases assessed for that criterion, with a Wilson 95\% confidence interval (CI).}
\label{tab:workspace-calibration}
\begin{tabularx}{\linewidth}{@{}>{\raggedright\arraybackslash}X r r l@{}}
\toprule
Assessment criterion & Cases & Rate (\%) & 95\% CI \\
\midrule
Agreement on recovery need & 96/100 & 96.0 & [90.2, 98.4] \\
Reconstruction success & 38/46 & 82.6 & [69.3, 90.9] \\
Success excluding binary limitations & 38/43 & 88.4 & [75.5, 94.9] \\
Unsupported file generation & 0/41 & 0.0 & [0.0, 8.6] \\
Inclusion of solution-created files & 1/41 & 2.4 & [0.4, 12.6] \\
\bottomrule
\end{tabularx}
\end{table}

\begin{table}[t]\centering\small
\caption{Workspace and environment recovery results. Workspace quality is assessed by human review; environment quality is assessed through the replay score, which measures agreement with the original tool-call results. Denominators indicate the cases assessed for each metric.}
\label{tab:rq1-combined}
\begin{tabular}{llrl}
\toprule Layer & Metric & Denominator & Result \\\midrule
Workspace & Need-recovery agreement & 100 & 96/100 (96.0\%) \\
 & Reconstruction success & 46 & 38/46 (82.6\%) \\
 & Success excluding binary limitations & 43 & 38/43 (88.4\%) \\
 & Unsupported file generation & 41 & 0/41 (0.0\%) \\
Environment & Successfully reconstructed and scored & 143 & 132/143 (92.3\%) \\
 & Mean replay score & 132 & 0.9175 \\
 & Replay score $\ge0.8$ & 132 & 119/132 (90.2\%) \\\bottomrule
\end{tabular}\end{table}

\subsubsection{Environment acceptance and repair studies}
\label{app:environment-calibration}
\paragraph{Check feasibility, deployment, and replay outcomes.}
We select a calibration set of 143 cases from the 14,603 coding traces to assess environment recovery, using the final replay score to evaluate the recovered workspace and environment jointly. After checking platform and hardware feasibility, we build and launch each supported environment with $\widehat W_0$ and replay the recorded tool calls. This yields 132 scored reconstructions, with ten cases excluded for unsupported platform or hardware requirements and one assessment incomplete.

Here, trajectory equivalence is quantified by the replay score in Eq.~\ref{eq:equiv-main}: a score of 1 over the complete tool-call sequence indicates that the recovered workspace and environment reproduce the original recorded results. Let $\sV$ index the 132 scored recovery cases and $v_j\in[0,1]$ the final replay score for case $j\in\sV$. We summarize both the average score and the fraction of cases reaching 0.8,
\begin{equation}
\bar v=\frac{1}{|\sV|}\sum_{j\in\sV}v_j=0.9175,
\qquad
\frac{|\{j\in\sV\mid v_j\ge0.8\}|}{|\sV|}
=\frac{119}{132}=90.2\%.
\label{eq:environment-audit}
\end{equation}
These results show that most recovered environments reproduce the recorded tool behavior closely.

\paragraph{Compare construction, launch repair, and full recovery.}
To assess the contributions of cold-start repair and replay-guided refinement, we compare the three conditions in Table~\ref{tab:rq1-env-ablation}. The launcher agent introduced in Appendix~\ref{app:environment-recovery} handles build and startup failures; the full workflow also uses temporal dependency inference and replay evaluation to guide environment tuning.

For the 29 cases evaluated under +Launcher, we define difficulty using the replay score $s$: Easy ($s\ge0.95$), Medium ($0.70\le s<0.95$), and Hard ($s<0.70$). Full recovery contributes its 26 available scored cases; differences compare condition-level means within these groups.

\begin{table}[t]\centering\small
\caption{Replay scores under three environment-recovery conditions. Single-shot makes one construction attempt, counting build failures as zero. +Launcher adds an agent that builds, cold-starts, and repairs the environment. Full also includes dependency inference and replay-guided refinement. Difference reports Full minus +Launcher.}
\label{tab:rq1-env-ablation}
\begin{tabular}{lrrrr}
\toprule Stratum & Single-shot & +Launcher & Full & Difference \\
\midrule
Hard & .0750 & .5303 & .8209 & +.2906 \\
Medium & .0000 & .8461 & .9165 & +.0704 \\
Easy & .0000 & .9874 & .9320 & -.0554 \\
\bottomrule
\end{tabular}\end{table}

Full recovery raises the reported mean from 0.5303 to 0.8209 in the hard stratum and from 0.8461 to 0.9165 in the medium stratum, while the easy-stratum mean falls from 0.9874 to 0.9320. This pattern locates the main benefit in cases where launch repair still leaves behavioral discrepancies, while the decline in easy cases shows that further repair can reduce replay agreement.

\paragraph{Assess replay-guided refinement.}
Within the environment-recovery calibration set, cases with both pre-refinement and final scores improve from a mean replay score of 0.8192 to 0.9175. This result supports replay-guided refinement as a means of improving recovery quality. Together, these studies assess deployment feasibility, replay fidelity, and the gains from refinement.

\subsubsection{Human and judge agent calibration}
\label{app:judge-calibration}

\begin{figure}[t]\centering
\includegraphics[width=.91\linewidth]{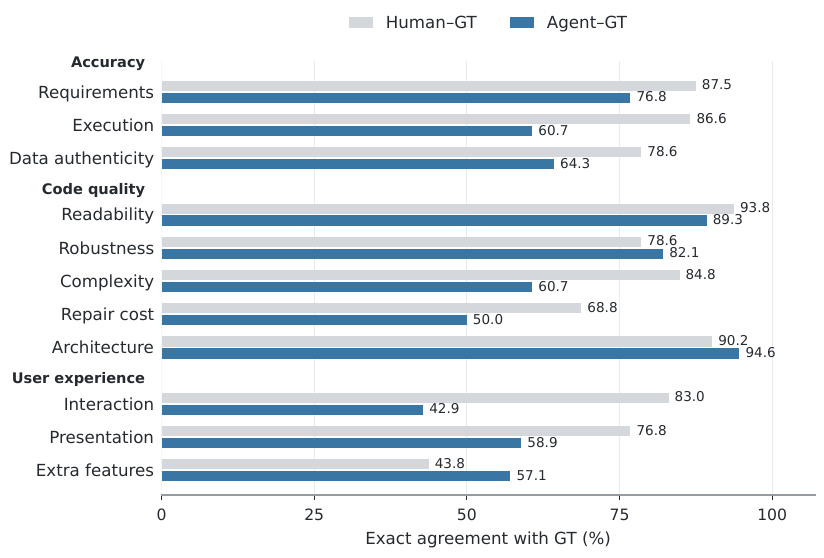}
\caption{Exact agreement with the adjudicated human reference (GT) across rubric dimensions. Human--GT averages the two annotators\textquotesingle{} agreement rates. Agent--GT compares the judge agent\textquotesingle{}s grade, obtained by rounding its mean over three scoring rounds, with the same reference.}
\label{fig:judge-calibration}
\end{figure}

The calibration study uses the same 56 generation artifacts and adjudicated GT to compare automated grades with two independent human annotators. Figure~\ref{fig:judge-calibration} reports dimension-level exact agreement and highlights differences between the judge agent and the human average. This comparison identifies strengths and weaknesses across rubric dimensions.

Agreement varies substantially across dimensions, with lower judge agent accuracy on several criteria requiring runtime or visual interpretation as well as some source-based criteria. These differences motivate complementing exact agreement with chance-corrected measures.

\subsubsection{Human--human, agent--human, and GT-referenced agreement}
\label{app:paired-judge-agreement}

GT comes from a third human annotation round: an annotator reviews Human A's and Human B's scores and rationales together with the artifact itself to assign the final reference grades (Appendix~\ref{app:annotation}).

\begin{table}[t]\centering\footnotesize
\setlength{\tabcolsep}{3pt}
\caption{Agreement between human annotators (H--H), between the judge agent and human annotators (A--H), and with the adjudicated human reference (GT). Cohen\textquotesingle{}s $\kappa$ measures agreement beyond chance. W-1 is the proportion of grades within one point of GT; bias is the mean signed difference from GT, with positive values indicating higher grades.}
\label{tab:rq2-kappa}
\begin{tabular}{lrrrrrrrr}\toprule
 & \multicolumn{2}{c}{Direct comparison} & \multicolumn{3}{c}{Human--GT} & \multicolumn{3}{c}{Agent--GT} \\
\cmidrule(lr){2-3}\cmidrule(lr){4-6}\cmidrule(lr){7-9}
Dimension & H--H $\kappa$ & A--H $\kappa$ & $\kappa$ & W-1 & Bias & $\kappa$ & W-1 & Bias \\
\midrule
\multicolumn{9}{l}{\textit{Accuracy}} \\
Requirement implementation & $0.303$ & $0.075$ & $0.532$ & $0.946$ & $-0.11$ & $0.188$ & $1.000$ & $-0.09$ \\
Code execution & $0.622$ & $0.283$ & $0.762$ & $0.938$ & $-0.04$ & $0.287$ & $0.946$ & $+0.16$ \\
Data authenticity & $0.312$ & $0.265$ & $0.589$ & $0.955$ & $-0.06$ & $0.248$ & $0.911$ & $+0.16$ \\
\textit{Weighted group mean} & $0.432$ & $0.187$ & $0.632$ & $0.944$ & $-0.07$ & $0.236$ & $0.965$ & $+0.05$ \\
\midrule
\multicolumn{9}{l}{\textit{Code quality}} \\
Readability & $0.000$ & $-0.028$ & $0.148$ & $1.000$ & $+0.03$ & $-0.057$ & $1.000$ & $+0.00$ \\
Robustness & $0.074$ & $0.000$ & $0.452$ & $0.991$ & $+0.01$ & $0.000$ & $1.000$ & $-0.18$ \\
Complexity & $0.407$ & $-0.008$ & $0.631$ & $1.000$ & $+0.06$ & $-0.020$ & $1.000$ & $+0.14$ \\
Repair cost & $0.160$ & $0.064$ & $0.425$ & $0.973$ & $-0.02$ & $0.013$ & $0.982$ & $+0.12$ \\
Architecture & $0.317$ & $0.306$ & $0.397$ & $0.982$ & $-0.07$ & $0.387$ & $1.000$ & $+0.03$ \\
\textit{Weighted group mean} & $0.170$ & $0.068$ & $0.397$ & $0.990$ & $-0.00$ & $0.070$ & $0.997$ & $-0.01$ \\
\midrule
\multicolumn{9}{l}{\textit{User experience}} \\
Interface and interaction & $0.527$ & $0.130$ & $0.732$ & $0.938$ & $-0.05$ & $0.138$ & $0.946$ & $+0.12$ \\
Information presentation & $0.366$ & $0.288$ & $0.578$ & $0.920$ & $-0.12$ & $0.324$ & $1.000$ & $-0.02$ \\
Feature richness & $0.289$ & $0.113$ & $0.123$ & $0.705$ & $-0.70$ & $0.135$ & $0.911$ & $-0.23$ \\
\textit{Weighted group mean} & $0.434$ & $0.180$ & $0.579$ & $0.893$ & $-0.18$ & $0.200$ & $0.958$ & $+0.02$ \\
\midrule
\textbf{Weighted macro mean} & $0.341$ & $0.144$ & $0.542$ & $0.952$ & $-0.06$ & $0.173$ & $0.975$ & $+0.02$ \\
\bottomrule\end{tabular}
\end{table}

Table~\ref{tab:rq2-kappa} distinguishes direct inter-rater agreement from agreement with adjudicated GT. Each automated grade is the rounded mean of three scoring rounds, yielding 616 artifact--dimension pairs. For each dimension, human--human $\kappa$ compares Human A with Human B; agent--human $\kappa$ averages the separate Agent--Human A and Agent--Human B values. Human--GT likewise averages the two annotators' separate comparisons with GT, while Agent--GT compares the agent directly with GT. Cohen's $\kappa=(p_{\mathrm{o}}-p_{\mathrm{e}})/(1-p_{\mathrm{e}})$ adjusts observed agreement $p_{\mathrm{o}}$ for the agreement $p_{\mathrm{e}}$ expected from the marginal grade distributions. Within-one agreement and signed bias provide complementary GT-referenced diagnostics.

The final weights follow expert priorities, are confirmed through reassessment, and are listed in Table~\ref{tab:rubric-dims} (Appendix~\ref{app:rubric-weights}). For a dimension-level agreement statistic $M_d$, the overall summary is $\sum_d w_d M_d$; a group summary normalizes the weights over that group's dimensions. Thus, the reported weighted $\kappa$ averages dimension-level $\kappa$ values, emphasizing agreement on the aspects of quality that experts prioritize for users.

\paragraph{Overall agreement.}
On individual criteria, the judge agent can exceed the human average in both exact agreement and $\kappa$, as observed for feature richness (Figure~\ref{fig:judge-calibration} and Table~\ref{tab:rq2-kappa}). These calibration results support partial replacement of manual annotation with calibrated agent scoring. Because our evaluation centers on user experience, human review remains important for nuanced, context-dependent quality judgments. Appendix~\ref{app:model-panel-configuration} specifies the judge agent used for the model panel.

\subsubsection{From recovered tasks to model rollout and artifact scoring}
\label{app:model-rollout}
This section details the execution and scoring protocol for the R1/R2 comparison in \S\ref{sec:rq3}, reported in Table~\ref{tab:rq3-panel} and Figure~\ref{fig:feedback-gains}.

\paragraph{Execute the request to produce the R1 artifact.}
Each task supplies the constructed user request $q$, recovered workspace $\widehat W_0$, and recovered initial environment $\widehat E_0$ defined in Appendix~\ref{app:recovery}. We package these inputs for the Terminal-Bench 2.0 execution infrastructure~\citep{terminalbench2026}, using our reconstructed tasks. Claude Code and OpenHands serve as two agent harnesses through which candidate models read files, edit code, and execute commands under the same evaluation protocol. Each initial attempt starts from a fresh copy of the task context, allowing the candidate to inspect the available inputs and develop its own solution. The execution infrastructure retains the resulting files and action record for evaluation. A small number of runs affected by API rate limits or environment failures are retried before their artifacts enter the same scoring and pairing procedure used for both harnesses.

\paragraph{Score the R1 artifact.}
Let $y$ denote the artifact produced by this execution, comprising the files created or modified to address $q$. For a software-development request, successful delivery means that the project can run and provide the requested behavior in the task environment. For a request targeting scripts, data, or other files, the delivered directory should contain the requested outputs and support the operations needed to verify them. The scoring workflow in Appendix~\ref{app:scoring} combines source inspection with execution or rendered evidence and applies the criteria in Appendix~\ref{app:rubric} to produce dimension scores, supporting rationales, and an aggregate artifact score.

\paragraph{Revise and rescore the artifact in R2.}
The R1 artifact and its scoring evidence feed the revision workflow in Appendix~\ref{app:feedback}, which turns identified weaknesses into an instruction. In R2, the same candidate model follows this instruction to revise its preserved R1 artifact. We score the revised artifact under the same rubric and aggregation rule and compare both rounds on matched tasks. This produces the paired scores in Table~\ref{tab:rq3-panel} and the improvement, tie, and regression outcomes in Table~\ref{tab:feedback-task-outcomes}, testing whether rubric-guided feedback improves the delivered artifact.

\subsubsection{Model-panel configuration and repeated sampling}
\label{app:model-panel-configuration}
The model comparison reports reassessed Total scores under the final expert-informed weighting scheme in Table~\ref{tab:rubric-dims}. Each artifact's score is a weighted mean over applicable dimensions, and the model Total averages these scores equally across tasks (Appendix~\ref{app:panel-scoring}). Task identities are matched to the canonical benchmark set $\sB$, and GLM-5.3-Flash supplies the judge agent for the model panel. Table~\ref{tab:panel-dimensions} expands the generation dimensions underlying Table~\ref{tab:rq3-panel}. Paired revision columns use the matched subsets defined in Appendix~\ref{app:feedback}.

\paragraph{Model references.}
Official model cards and release documentation describe DeepSeek-V4-Pro~\citep{deepseek_v4_modelcard_2026}, GLM-5.2 and GLM-5.3~\citep{zai_glm52_modelcard_2026,zai_glm53_modelcard_2026}, GPT-5.5 and GPT-5.6 sol~\citep{openai_gpt55_modelcard_2026,openai_gpt56_2026}, Kimi-K3~\citep{moonshot_k3_modelcard_2026}, MiniMax-M3~\citep{minimax_m3_modelcard_2026}, LongCat-2.0~\citep{longcat_20_modelcard_2026}, Fable 5~\citep{anthropic_fable5_2026}, and Gemini 3.6 Flash~\citep{google_gemini36_modelcard_2026}. Table~\ref{tab:rq3-panel} links each model name directly to its documentation.

\paragraph{Gemini compatibility with the Claude Code harness.}
\label{app:gemini-compatibility}
In our Claude Code deployment, the Anthropic-compatible gateway does not support Gemini's \texttt{thought\_signature} metadata, a limitation identified during Gemini 3.6 Flash integration for multimodal evaluation. Gemini uses these signatures to preserve reasoning context across tool calls and requires them to be returned in subsequent requests~\citep{google_thought_signatures2026}. We therefore exclude Gemini--Claude Code results from the reported comparisons to avoid confounding artifact quality with this protocol incompatibility. Gemini--OpenHands results remain included.

\begin{table}[t]
\centering\scriptsize
\setlength{\tabcolsep}{2.3pt}
\renewcommand{\arraystretch}{1.06}
\caption{Dimension-level breakdown of the Table~\ref{tab:rq3-panel} results under Claude Code.}
\label{tab:panel-dimensions}
\resizebox{\linewidth}{!}{%
\begin{tabular}{ll*{11}{r}}
\toprule
Model & Round & \multicolumn{3}{c}{Accuracy} & \multicolumn{5}{c}{Code quality} & \multicolumn{3}{c}{User experience} \\
\cmidrule(lr){3-5}\cmidrule(lr){6-10}\cmidrule(lr){11-13}
 & & Req. & Exec. & Data & Read. & Rob. & Complex. & Repair & Arch. & Interact. & Present. & Richness \\
\midrule
DeepSeek & R1 & 76.2 & 74.2 & 74.9 & 83.3 & 81.1 & 82.9 & 81.5 & 83.1 & 64.8 & 66.4 & 67.6 \\
 & R2 & 98.3 & 96.3 & 97.0 & 94.1 & 91.9 & 93.7 & 92.3 & 93.8 & 88.0 & 89.6 & 90.8 \\
GLM-5.2 & R1 & 80.9 & 78.9 & 79.6 & 87.8 & 85.6 & 87.4 & 86.0 & 87.6 & 66.9 & 68.5 & 69.7 \\
 & R2 & 95.9 & 93.9 & 94.6 & 97.4 & 95.2 & 97.0 & 95.6 & 97.1 & 86.5 & 88.1 & 89.3 \\
GLM-5.3 & R1 & 84.6 & 82.6 & 83.3 & 91.6 & 89.4 & 91.2 & 89.8 & 91.4 & 68.6 & 70.2 & 71.4 \\
 & R2 & 98.2 & 96.2 & 96.9 & 100.0 & 98.1 & 99.6 & 98.4 & 99.8 & 92.0 & 93.6 & 94.8 \\
GPT-5.5 & R1 & 77.9 & 75.9 & 76.6 & 87.6 & 85.4 & 87.2 & 85.8 & 87.4 & 57.9 & 59.5 & 60.7 \\
 & R2 & 90.2 & 88.2 & 88.8 & 96.8 & 94.6 & 96.4 & 95.0 & 96.5 & 76.8 & 78.4 & 79.6 \\
GPT-5.6 sol & R1 & 87.0 & 85.0 & 85.7 & 93.4 & 91.2 & 93.0 & 91.6 & 93.1 & 73.4 & 75.0 & 76.2 \\
 & R2 & 95.5 & 93.5 & 94.1 & 98.3 & 96.1 & 97.9 & 96.5 & 98.0 & 87.0 & 88.6 & 89.8 \\
Kimi-K3 & R1 & 79.5 & 77.5 & 78.2 & 88.9 & 86.7 & 88.5 & 87.1 & 88.6 & 58.8 & 60.4 & 61.6 \\
 & R2 & 89.4 & 87.4 & 88.0 & 94.8 & 92.6 & 94.4 & 93.0 & 94.5 & 75.0 & 76.6 & 77.8 \\
MiniMax-M3 & R1 & 76.9 & 74.9 & 75.6 & 86.3 & 84.1 & 85.9 & 84.5 & 86.1 & 57.4 & 59.0 & 60.2 \\
 & R2 & 92.0 & 90.0 & 90.6 & 97.2 & 95.0 & 96.8 & 95.4 & 96.9 & 80.7 & 82.3 & 83.5 \\
LongCat & R1 & 83.3 & 81.3 & 82.0 & 91.8 & 89.6 & 91.4 & 90.0 & 91.5 & 62.9 & 64.5 & 65.7 \\
 & R2 & 92.1 & 90.1 & 90.7 & 97.1 & 94.9 & 96.7 & 95.3 & 96.8 & 78.3 & 79.9 & 81.1 \\
Fable 5 & R1 & 93.4 & 91.4 & 92.1 & 97.0 & 94.8 & 96.6 & 95.2 & 96.7 & 80.3 & 81.9 & 83.1 \\
 & R2 & 98.4 & 96.4 & 97.0 & 100.0 & 98.3 & 99.7 & 98.6 & 99.8 & 89.5 & 91.1 & 92.3 \\
\bottomrule
\end{tabular}%
}
\end{table}

\begin{table}[t]
\centering\scriptsize
\setlength{\tabcolsep}{2.3pt}
\renewcommand{\arraystretch}{1.06}
\caption{Dimension-level breakdown of the Table~\ref{tab:rq3-panel} results under OpenHands.}
\label{tab:panel-dimensions-openhands}
\resizebox{\linewidth}{!}{%
\begin{tabular}{ll*{11}{r}}
\toprule
Model & Round & \multicolumn{3}{c}{Accuracy} & \multicolumn{5}{c}{Code quality} & \multicolumn{3}{c}{User experience} \\
\cmidrule(lr){3-5}\cmidrule(lr){6-10}\cmidrule(lr){11-13}
 & & Req. & Exec. & Data & Read. & Rob. & Complex. & Repair & Arch. & Interact. & Present. & Richness \\
\midrule
DeepSeek & R1 & 89.0 & 87.0 & 87.6 & 92.9 & 90.7 & 92.5 & 91.1 & 92.7 & 73.5 & 75.1 & 76.3 \\
 & R2 & 93.2 & 91.2 & 91.9 & 98.5 & 96.3 & 98.1 & 96.7 & 98.3 & 88.4 & 90.0 & 91.2 \\
Gemini 3.6 & R1 & 83.7 & 81.7 & 82.3 & 91.1 & 88.9 & 90.7 & 89.3 & 90.8 & 69.2 & 70.8 & 72.0 \\
 & R2 & 92.1 & 90.1 & 90.8 & 97.0 & 94.8 & 96.6 & 95.2 & 96.8 & 84.8 & 86.4 & 87.6 \\
GLM-5.2 & R1 & 88.0 & 86.0 & 86.7 & 94.0 & 91.8 & 93.6 & 92.2 & 93.7 & 66.1 & 67.7 & 68.9 \\
 & R2 & 95.3 & 93.3 & 94.0 & 99.6 & 97.4 & 99.2 & 97.8 & 99.3 & 87.1 & 88.7 & 89.9 \\
GLM-5.3 & R1 & 83.2 & 81.2 & 81.9 & 91.1 & 88.9 & 90.7 & 89.3 & 90.9 & 70.6 & 72.2 & 73.4 \\
 & R2 & 91.5 & 89.5 & 90.2 & 96.0 & 93.8 & 95.6 & 94.2 & 95.8 & 77.1 & 78.7 & 79.9 \\
GPT-5.5 & R1 & 85.2 & 83.2 & 83.9 & 91.1 & 88.9 & 90.7 & 89.3 & 90.8 & 64.0 & 65.6 & 66.8 \\
 & R2 & 94.4 & 92.4 & 93.1 & 97.0 & 94.8 & 96.6 & 95.2 & 96.7 & 86.9 & 88.5 & 89.7 \\
GPT-5.6 sol & R1 & 89.3 & 87.3 & 88.0 & 94.5 & 92.3 & 94.1 & 92.7 & 94.2 & 76.7 & 78.3 & 79.5 \\
 & R2 & 96.5 & 94.5 & 95.1 & 99.3 & 97.1 & 98.9 & 97.5 & 99.0 & 91.9 & 93.5 & 94.7 \\
Kimi-K3 & R1 & 78.2 & 76.2 & 76.8 & 87.8 & 85.6 & 87.4 & 86.0 & 87.5 & 58.0 & 59.6 & 60.8 \\
 & R2 & 87.8 & 85.8 & 86.4 & 93.7 & 91.5 & 93.3 & 91.9 & 93.4 & 74.8 & 76.4 & 77.6 \\
MiniMax-M3 & R1 & 85.4 & 83.4 & 84.0 & 91.6 & 89.4 & 91.2 & 89.8 & 91.3 & 64.4 & 66.0 & 67.2 \\
 & R2 & 91.2 & 89.2 & 89.9 & 99.0 & 96.8 & 98.6 & 97.2 & 98.7 & 84.6 & 86.2 & 87.4 \\
LongCat & R1 & 79.8 & 77.8 & 78.5 & 89.1 & 86.9 & 88.7 & 87.3 & 88.9 & 60.0 & 61.6 & 62.8 \\
 & R2 & 89.3 & 87.3 & 88.0 & 94.9 & 92.7 & 94.5 & 93.1 & 94.6 & 76.9 & 78.5 & 79.7 \\
Fable 5 & R1 & 90.0 & 88.0 & 88.6 & 94.0 & 91.8 & 93.6 & 92.2 & 93.7 & 75.1 & 76.7 & 77.9 \\
 & R2 & 96.2 & 94.2 & 94.9 & 98.5 & 96.3 & 98.1 & 96.7 & 98.2 & 91.6 & 93.2 & 94.4 \\
\bottomrule
\end{tabular}%
}
\end{table}

\paragraph{Aggregate repeated samples for cost and quality comparison.}
\label{app:cost-effectiveness}
Experiments are distributed across multiple groups of compute nodes, and their outputs are consolidated by benchmark task identity to complete coverage of the original 129-task \method benchmark. We collect three samples for each configuration and assess variability across these samples, applying the same task-specific scoring and aggregation rules throughout. Matching records by task identity provides a common basis for comparing model configurations across distributed runs. Figure~\ref{fig:cost-effectiveness} presents the first-round comparison under Claude Code, relating artifact quality to candidate-model execution cost. This joint view complements the capability-level scores by showing how the quality of delivered artifacts relates to the resources used to produce them.

\subsubsection{Feedback ablation: an unguided second development pass}
\label{app:feedback-ablation}
\paragraph{Hold the starting artifact fixed.}
To distinguish the benefit of rubric-derived feedback from another development opportunity, we compare two second-round conditions from the same R1 artifact and recovered environment. The feedback condition follows Appendix~\ref{app:feedback}; the unguided condition receives the original user request verbatim, without weaknesses, scoring rationales, or additional improvement instructions. The candidate model, recorded sampling settings, and execution limits are matched between the two conditions. We compare the latest available attempt per task and retain the common set with verified agent participation and complete three-rollout scoring in all three conditions.

\paragraph{Compare Total scores on the same tasks.}
Table~\ref{tab:feedback-ablation} reports the detailed values underlying Figure~\ref{fig:feedback-ablation}. Each row lists the R1 baseline, the unguided second pass (U2), the feedback-guided second pass (R2), and the corresponding gains over R1.

\begin{table}[!htbp]
\centering\small
\setlength{\tabcolsep}{4.5pt}
\renewcommand{\arraystretch}{1.08}
\caption{Detailed values for Figure~\ref{fig:feedback-ablation}. R1 denotes initial generation; R2 is revision with rubric feedback, and U2 is revision using the unchanged original request. Both revisions start from the same R1 artifact and environment. Scores are \% of maximum; $\Delta$R2 and $\Delta$U2 measure gains over R1, and all differences are in pp.}
\label{tab:feedback-ablation}
\begin{tabular}{l*{6}{r}}
\toprule
Model & R1 & R2 & U2 & $\Delta$R2 & $\Delta$U2 & R2$-$U2 \\
\midrule
\multicolumn{7}{l}{\textbf{Claude Code}} \\
DeepSeek-V4-Pro & 76.85 & 96.13 & 82.63 & $+19.28$ & $+5.78$ & $+13.50$ \\
GLM-5.2 & 80.96 & 94.13 & 86.39 & $+13.17$ & $+5.43$ & $+7.74$ \\
GLM-5.3 & 84.76 & 97.29 & 90.20 & $+12.53$ & $+5.44$ & $+7.09$ \\
GPT-5.5 & 78.10 & 89.88 & 82.85 & $+11.78$ & $+4.75$ & $+7.03$ \\
GPT-5.6 sol & 86.23 & 94.87 & 89.71 & $+8.64$ & $+3.48$ & $+5.16$ \\
Kimi-K3 & 78.93 & 88.07 & 82.62 & $+9.14$ & $+3.69$ & $+5.45$ \\
MiniMax-M3 & 76.64 & 91.57 & 83.61 & $+14.93$ & $+6.97$ & $+7.96$ \\
LongCat-2.0 & 82.65 & 92.04 & 86.44 & $+9.39$ & $+3.79$ & $+5.60$ \\
Fable 5 & 92.42 & 97.37 & 94.42 & $+4.95$ & $+2.00$ & $+2.95$ \\
\midrule
\multicolumn{7}{l}{\textbf{OpenHands}} \\
DeepSeek-V4-Pro & 85.72 & 93.75 & 90.79 & $+8.03$ & $+5.07$ & $+2.96$ \\
Gemini 3.6 Flash & 82.89 & 91.76 & 85.27 & $+8.87$ & $+2.38$ & $+6.49$ \\
GLM-5.2 & 84.27 & 95.09 & 88.75 & $+10.82$ & $+4.48$ & $+6.34$ \\
GLM-5.3 & 84.90 & 90.41 & 86.79 & $+5.51$ & $+1.89$ & $+3.62$ \\
GPT-5.5 & 82.86 & 94.62 & 88.19 & $+11.76$ & $+5.33$ & $+6.43$ \\
GPT-5.6 sol & 87.46 & 95.89 & 91.28 & $+8.43$ & $+3.82$ & $+4.61$ \\
Kimi-K3 & 77.54 & 86.16 & 81.44 & $+8.62$ & $+3.90$ & $+4.72$ \\
MiniMax-M3 & 82.40 & 91.97 & 88.21 & $+9.57$ & $+5.81$ & $+3.76$ \\
LongCat-2.0 & 79.52 & 88.18 & 83.44 & $+8.66$ & $+3.92$ & $+4.74$ \\
Fable 5 & 88.98 & 95.91 & 92.12 & $+6.93$ & $+3.14$ & $+3.79$ \\
\bottomrule
\end{tabular}
\end{table}

\paragraph{Feedback adds value beyond another development pass.}
An unguided second pass improves the mean in every reported configuration, confirming that an additional development opportunity already has value. Rubric-guided revision achieves a higher mean than the unguided pass across the detailed comparisons in Table~\ref{tab:feedback-ablation}.

These results complement the qualitative cases in Appendix~\ref{app:feedback-qualitative}: the feedback identifies specific defects that a repeated original request leaves implicit.

\subsection{Rubric-Guided Revision and Analysis}
\label{app:feedback}

\subsubsection{Revision Protocol}

This workflow provides a practical test of the rubric by examining whether its diagnoses guide artifact improvement (\S\ref{sec:rq3}). R1 is the candidate's initial artifact from the original request; R2 is its feedback-guided revision by the same model and harness in the preserved task environment. Both use the same rubric. A separate unguided revision starts from the same R1 artifact using only the original request, allowing the ablation to examine the contribution of targeted guidance beyond another development pass (Appendix~\ref{app:feedback-ablation}).

\begin{figure}[ht]\centering
\includegraphics[width=\linewidth]{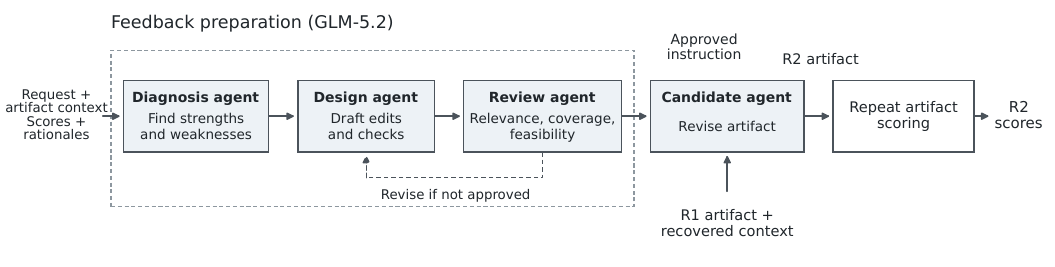}
\caption{Rubric-guided feedback and artifact revision. Diagnosis and design agents identify weaknesses and propose edits; a review agent checks the instruction and returns it for revision when needed. The candidate agent applies the approved instruction to its R1 artifact. The revised artifact is scored using the same rubric and the workflow in Figure~\ref{fig:judge-workflow}.}
\label{fig:feedback-workflow}
\end{figure}

\paragraph{Turn observed weaknesses into actionable guidance.}
As shown in Figure~\ref{fig:feedback-workflow}, feedback preparation uses the original request, artifact description, dimension scores, and supporting rationales. A diagnosis agent identifies evidence-backed weaknesses and strengths. A design agent translates the weaknesses into concrete actions, such as repairing a failed operation or completing a missing workflow, while stating which working behavior should be preserved. A review agent checks that the instruction remains faithful to the request, covers the identified priorities, and asks for feasible changes. Instructions that fail this check are revised before delivery. GLM-5.2 performs these preparation roles, and the candidate model carries out the resulting instruction.

The instruction describes what to inspect, the intended improvement, and how to verify it. It directs the candidate toward observed defects while leaving implementation choices to that candidate. The revision starts from its first-round artifact in the preserved task environment; reference implementations and adjudicated human scores are not supplied. This preserves the connection between a diagnosed weakness and the artifact being improved.

\paragraph{Compare each revision with its actual starting point.}
For task $i$ and model $m$, let $s^{(1)}_{i,m}$ and $s^{(2)}_{i,m}$ denote normalized R1 and R2 scores under the same rubric and aggregation rule (Appendix~\ref{app:scoring}). Under both Claude Code and OpenHands, each R2 artifact is paired with the R1 artifact used to generate its feedback. Let $\sP^{\mathrm{rev}}_m$ contain the indices of matched tasks with preserved starting artifacts, completed feedback preparation, candidate participation, and complete paired scoring. For $|\sP^{\mathrm{rev}}_m|>0$, we compute per-task differences and express the mean gain in percentage points (pp):
\begin{equation}
\Delta_{i,m}=s^{(2)}_{i,m}-s^{(1)}_{i,m},
\qquad
G_m=\frac{100}{|\sP^{\mathrm{rev}}_m|}\sum_{i\in\sP^{\mathrm{rev}}_m}\Delta_{i,m}.
\label{eq:feedback-gain}
\end{equation}
The sign of $\Delta_{i,m}$ identifies improvement, a tie, or regression. Means and outcome proportions use the same matched tasks. Under both the Claude Code and OpenHands harnesses, paired scoring requires scores for all applicable dimensions from three judge agent rollouts for each artifact in R1 and R2. We obtain 95\% confidence intervals from the 2.5th and 97.5th percentiles of mean changes across 10,000 bootstrap resamples of matched task pairs.

\paragraph{Interpret the changes through the rubric.}
Table~\ref{tab:rq3-panel} and Figure~\ref{fig:feedback-gains} summarize paired score changes; Table~\ref{tab:feedback-task-outcomes} reports task outcomes. The unguided comparison in Appendix~\ref{app:feedback-ablation} evaluates the contribution of this guidance beyond a second development pass.

\subsubsection{Qualitative study of feedback-guided revision}
\label{app:feedback-qualitative}
We inspect matched R1/R2 artifacts together with their feedback instructions, source changes, and recorded execution evidence. The examples below illustrate how a diagnosed weakness becomes an edit and how fully that edit improves the delivered workflow.

\paragraph{Completing an interface makes existing functionality accessible.}
A DeepSeek-V4-Pro workflow application initially contains backend APIs and client-side calls, but its front end remains a placeholder without routed pages or usable controls. Feedback identifies this missing layer and asks the model to expose the existing functionality through forms, navigation, and data views. The revision adds a router and pages for authentication, workflow management, monitoring, and task history. Recorded browser interactions confirm page rendering, form entry, and navigation. Interaction and information-presentation scores both rise from 0 to 2 across the three scoring rounds. The improvement comes from connecting existing implementation work to a visible, operable interface.

\paragraph{Repairing a blocking error unlocks executable behavior.}
A GLM-5.2 service cannot compile because of truncated source and inconsistent interfaces. Feedback directs the model to repair these defects and verify the login endpoint; R2 compiles and returns the requested JSON response. Appendix~\ref{app:feedback-case-study} presents the original request, feedback, source changes, and the corresponding artifact-score increase of 51.52~pp.

\paragraph{Connecting state, persistence, and controls improves data authenticity.}
A GLM-5.3 finance application already renders polished pages, but hard-coded summaries and an ineffective expense form prevent meaningful data updates. Feedback asks for a shared data layer and explicit connections between form controls, stored records, and derived totals. The revision introduces shared state with persistent storage and binds expense submission to the data update function. Source inspection confirms these connections, and the execution evidence records working navigation and form interactions. Here, feedback targets the data flow beneath an existing interface rather than its appearance.

\paragraph{Cross-component consistency determines whether a repair is complete.}
A MiniMax-M3 commerce application receives feedback to implement missing backend endpoints and replace placeholder dependencies. Its revision adds product, group-purchase, cart, and order controllers, and endpoint probes return structured business data. However, frontend requests and routes remain inconsistent with the backend: the initial page loads, but navigation and data retrieval still fail. The code-execution score remains at 1 in all three R2 scoring rounds. Completing components individually therefore does not ensure a usable end-to-end workflow.

Together, these cases locate the clearest observed benefits in concrete, verifiable defects. They also explain why models can show different gains: an initially blocked or incomplete artifact offers substantial room for improvement, while realizing that gain requires the model to carry the repair through the relevant files, dependencies, and interfaces. The outcome depends on both the initial defect and the completeness of the implemented revision.

\subsubsection{Case Study: From a Failed Build to a Working Login Service}
\label{app:feedback-case-study}

This case follows GLM-5.2 under Claude Code through the feedback-guided revision protocol in \S\ref{sec:rq3}. The task is to add a callable JSON login endpoint to an existing Scala/Pekko HTTP application while preserving its DTOs, actor routing, and Spring integration. It therefore requires both local controller code and integration with a working service.

To develop this login service, GLM-5.2 first executes the original request in the recovered workspace and environment, producing the initial artifact R1. The judge agent evaluates this delivery, and the pipeline uses its scores and rationales to formulate a revision instruction. GLM-5.2 then applies that instruction to the preserved R1 artifact in the same task environment, producing the revised artifact R2. Both artifacts address the same request and are assessed with the same rubric. In this case, R1 supplies much of the structure but cannot run; feedback-guided R2 raises the artifact's Total score from 46.97 to 98.49 (+51.52~pp).

\paragraph{The actual two-round task instructions.}
The \href{https://anonymous.4open.science/r/TRACE2EVAL-ICLR/prompts/login_service/README.md}{complete R1 and R2 instructions} are provided in the anonymous repository, with the Chinese portions translated into English. Filenames, line references, and task-directory paths are retained.

\paragraph{R1 implements the interface but stops short of execution.}
The candidate uses \texttt{Read} to inspect the User DTO and Maven configuration, \texttt{Edit} to add JSON-routing dependencies, and \texttt{Write} to create the authentication controller and HTTP bootstrap. Its subsequent \texttt{Bash} compilation attempt encounters missing Maven and Java executables. It searches for the tools but finishes without installing them or verifying a running endpoint. The later scoring run, with the build tools available, exposes truncated source and incompatible actor interfaces. Thus, R1's favorable readability and complexity grades coexist with an execution grade of zero: the controller is present, but the requested behavior remains unusable.

\paragraph{Feedback changes the candidate's work sequence.}
The R2 trajectory begins by reading the named files and using \texttt{TaskCreate} to turn the feedback into repair and verification tasks. The candidate again encounters the missing toolchain, but now installs OpenJDK and Maven through \texttt{Bash} before compiling. Feedback supplies a concrete completion condition, a service that builds and answers login requests, and the observed work proceeds toward that condition.

Source inspection also guides how the instruction is implemented. The candidate searches the room-factory call sites, aligns the factory signature with the actual caller, and replaces the incomplete HTTP helper with the JDK HTTP client. After removing the old HTTP dependency, compilation reports unresolved HTTP references in two dependent files; the candidate updates those callers and recompiles. Later stability edits introduce two further type errors, which it also fixes before continuing. The resulting sequence combines rubric-derived repair priorities with fresh compiler feedback, carrying changes through the affected components.

\paragraph{Selected tool interactions expose the improvement mechanism.}
The following sequence comes from the candidate's own tool calls and returned observations, rather than its final self-report. Calls are abridged to show the relevant arguments; file paths are expressed relative to the task workspace.
\begin{enumerate}[leftmargin=1.6em,itemsep=2pt,topsep=2pt,before=\raggedright]
\item \textbf{Inspect and localize.} \texttt{Read(HttpService.scala)}, \texttt{Read(RoomFactory.scala)}, and \texttt{Read(OutboundActor.scala)} inspect the feedback's named defects. A \texttt{Bash} search for \texttt{RoomFactory.createRoom} finds the actual two-user call, allowing a concrete interface repair.
\item \textbf{Enable executable checks.} \texttt{Bash(apt-get install ... openjdk-21-jdk-headless maven)} supplies the missing toolchain. The candidate can now obtain compiler diagnostics and use them to direct subsequent edits.
\item \textbf{Close the repair loop.} \texttt{Write} and \texttt{Edit} update the HTTP helper, its callers, and actor handlers. Repeated \texttt{Bash(mvn compile)} calls expose remaining type errors; after the follow-up edits, the build succeeds. A final \texttt{Bash(mvn clean compile)} also returns \texttt{BUILD SUCCESS}.
\item \textbf{Verify the requested behavior.} After launching the application, \texttt{Bash(curl ... /auth/ping)} returns a healthy status and \texttt{Bash(curl -X POST ... /auth/login)} returns the user's fields and session information. A subsequent check parses the response and confirms that \texttt{accessToken} differs from \texttt{sessionId}, directly testing another feedback item.
\item \textbf{Make the result inspectable.} \texttt{Write} adds a Vue login form, followed by a successful \texttt{Bash(npx vite build)}. When the development server hits a file-watcher limit, the candidate switches to build preview and captures the rendered page. The later scoring runs confirm the backend response and assess the added interface.
\end{enumerate}

\begin{table}[htbp]
\centering\small
\caption{Artifact quality before and after feedback-guided revision in the GLM-5.2 login-service case under Claude Code. Dimension scores average three judge agent rounds on a 0--2 scale. For this illustrative case, Total is the unweighted mean of the eleven dimensions, rescaled to 0--100. Changes are R2 minus R1, in rubric points for dimensions and pp for Total.}
\label{tab:feedback-case-study}
\begin{tabular}{lrrr}
\toprule
Rubric dimension & R1 & R2 & Change \\
\midrule
Requirement implementation & 1.00 & 2.00 & +1.00 \\
Code execution & 0.00 & 2.00 & +2.00 \\
Data authenticity & 1.00 & 2.00 & +1.00 \\
Readability & 2.00 & 2.00 & 0.00 \\
Robustness & 1.00 & 2.00 & +1.00 \\
Computational complexity & 2.00 & 2.00 & 0.00 \\
Repair cost & 1.33 & 2.00 & +0.67 \\
Architecture & 1.67 & 2.00 & +0.33 \\
Interface and interaction & 0.00 & 2.00 & +2.00 \\
Information presentation & 0.00 & 2.00 & +2.00 \\
Feature richness & 0.33 & 1.67 & +1.33 \\
\midrule
Total (\%) & 46.97 & 98.49 & +51.52 \\
\bottomrule
\end{tabular}
\end{table}

\paragraph{From targeted feedback to measured gains.}
Table~\ref{tab:feedback-case-study} shows the largest dimension-level gains in code execution, interface and interaction, and information presentation, each rising from 0 to 2, while readability and complexity retain their maximal R1 grades. The trajectory explains this pattern: explicit defect locations focus inspection, preservation instructions retain useful structure, and required execution checks expose incomplete repairs. Toolchain availability and cross-file dependencies influence progress, while compiler diagnostics and endpoint responses provide immediate feedback for the next action. Removing the build blocker makes existing functionality usable; the minimal interface then makes that functionality visible to the user. This case illustrates how the rubric can turn quality judgments into concrete repair and verification steps, complementing the aggregate feedback ablation in Appendix~\ref{app:feedback-ablation}.

\subsection{Profile Distribution Summary}
\label{app:full-profile-tables}

The distribution analysis in Appendix~\ref{app:data-construction} characterizes online requests by their intended activity, application setting, technical work, and user role. These four complementary aspects define the profile taxonomy used for task selection in \S\ref{sec:trace-retrieval}. Task type and technical domain each have broad and fine-grained levels, giving six axes with 171 category entries in total.

\paragraph{Broad task type (\texttt{x1}).}
The profile begins with what the user asks the agent to do. Axis \texttt{x1} groups the overall activity into ten category entries, providing a first-level account of the mix of requested work in online traffic.

\paragraph{Fine-grained task type (\texttt{x2}).}
Within these broad activity families, \texttt{x2} specifies the requested operation through 37 finer entries. This second level distinguishes specific task demands while retaining their connection to the broader activities represented by \texttt{x1}.

\paragraph{Usage scenario (\texttt{y}).}
The requested activity also has a context of use, captured by the usage-scenario axis \texttt{y}. Its 46 entries describe the application or business setting in which the result will be used, connecting the requested work to its practical purpose.

\paragraph{Broad technical domain (\texttt{z1}).}
Alongside this application context, \texttt{z1} identifies the broad technical field involved in fulfilling the request. Its ten entries characterize the knowledge and practices needed to complete the work.

\paragraph{Fine-grained technical domain (\texttt{z2}).}
These broad domains are further differentiated by \texttt{z2}, whose 51 entries capture more specific technical specializations. This second level preserves differences in the technical demands of requests that share the same \texttt{z1} label.

\paragraph{User role (\texttt{w}).}
The user-role axis \texttt{w} completes the profile by describing who requests the work and the capacity in which the result is needed. Its 17 entries capture the user's professional context. Combined with the activity, scenario, and technical-domain labels, this information allows the distribution comparisons to account for both the requested work and its intended use.

Table~\ref{tab:taxonomy-standard} lists the complete category taxonomy. The \href{https://anonymous.4open.science/r/TRACE2EVAL-ICLR/profiles/README.md}{anonymous repository} provides the corresponding online-traffic and SWE-bench distribution tables. Appendix~\ref{app:taxonomy-tables} explains how the labels and reporting denominators are used in the distribution comparisons.

{\footnotesize\begin{longtable}{>{\RaggedRight\arraybackslash}p{0.26\linewidth} >{\RaggedRight\arraybackslash}p{0.50\linewidth} r}
\caption{User-profile taxonomy derived from online coding-agent traffic. The six profile axes contain 171 categories; \#Cats gives the number of categories on each axis. Category names are translated from Chinese.}\label{tab:taxonomy-standard}\\
\toprule
Axis & Category & \#Cats \\ \midrule
\endfirsthead
\toprule Axis & Category & \#Cats \\ \midrule \endhead
\bottomrule \endfoot
\texttt{x1} - Task type L1 & Task Execution & \multirow{10}{*}{10} \\
 & Programming \& Technology & \\
 & Information Retrieval & \\
 & Data Processing & \\
 & Analysis \& Reasoning & \\
 & Content Creation & \\
 & Others & \\
 & Practical Guidance & \\
 & N/A & \\
 & Multimedia Processing & \\
\midrule
\texttt{x2} - Task type L2 & Automation Orchestration \& Monitoring & \multirow{37}{*}{37} \\
 & FI Feature Implementation: build a new capability from scratch & \\
 & CU Code Understanding: read, trace, or explain existing code & \\
 & General Knowledge Query & \\
 & BF Bug Fixing: diagnose and repair a defect & \\
 & Real-world Task Agent & \\
 & Anomaly Diagnosis \& Log Tracing & \\
 & FE Feature Enhancement: extend or improve an existing feature & \\
 & Platform \& Business Consulting & \\
 & Others & \\
 & CD Configuration \& Deployment: environment config, build, or deploy & \\
 & Timely News Query & \\
 & Functional Verification \& Acceptance & \\
 & Data Analysis \& Insight & \\
 & Runtime Env Config \& Self-healing & \\
 & Content Review \& Evaluation & \\
 & Data Extraction \& Cleaning & \\
 & Functional Writing & \\
 & Information Comparison \& Recommendation & \\
 & RF Refactoring: restructure code without changing behavior & \\
 & Logical Reasoning \& Computation & \\
 & Operational Advice / Guide & \\
 & TG Test Case Generation: write or extend test cases & \\
 & N/A & \\
 & Text Translation & \\
 & Conversational Text & \\
 & Data Labeling \& Classification & \\
 & Text Editing \& Optimization & \\
 & Planning \& Decision-making & \\
 & PO Performance Optimization: improve speed, memory, or efficiency & \\
 & Skill Acquisition \& Enhancement & \\
 & Creative Writing & \\
 & Image Generation \& Processing & \\
 & Tutoring \& Teaching & \\
 & Audio / Video Processing & \\
 & Cross-platform Auth \& Identity Mgmt & \\
 & Others & \\
\midrule
\texttt{y} - Usage scenario & Programming & \multirow{46}{*}{46} \\
 & Computers \& Electronics & \\
 & Finance & \\
 & Jobs \& Education & \\
 & Business \& Industrial & \\
 & Games & \\
 & Health -- Others & \\
 & Health -- Medical Consultation & \\
 & Health -- Medical / Rehab Services & \\
 & Science & \\
 & Home \& Garden -- Housekeeping \& Repair & \\
 & Home \& Garden -- Life Services & \\
 & Home \& Garden -- Others & \\
 & Arts \& Ent. -- Immersive \& Board Games & \\
 & Arts \& Ent. -- Wellness \& Bath & \\
 & Arts \& Ent. -- Others & \\
 & Arts \& Ent. -- Indoor Shows \& Ent. & \\
 & Autos \& Vehicles & \\
 & Food \& Drink -- Others & \\
 & Food \& Drink -- Meals \& Specialties & \\
 & Food \& Drink -- Takeout & \\
 & Food \& Drink -- Fresh \& Ingredients & \\
 & Law \& Government & \\
 & Real Estate & \\
 & Sports & \\
 & Shopping -- Group Buying & \\
 & Shopping -- Flash / Instant Retail & \\
 & Shopping -- Others & \\
 & Travel \& Trans. -- Tourism \& Vacation & \\
 & Travel \& Trans. -- Transit \& Ride-sharing & \\
 & Travel \& Trans. -- Lodging \& Hotels & \\
 & Travel \& Trans. -- Others & \\
 & Online Communities & \\
 & Books \& Literature & \\
 & Beauty \& Fitness -- Sports \& Fitness & \\
 & Beauty \& Fitness -- Medical Beauty & \\
 & Beauty \& Fitness -- Salon \& Hair & \\
 & Beauty \& Fitness -- Others & \\
 & Reference & \\
 & News & \\
 & Pets \& Animals & \\
 & Hobbies \& Leisure & \\
 & Internet \& Telecom & \\
 & People \& Society & \\
 & Others & \\
 & N/A & \\
\midrule
\texttt{z1} - Technical domain L1 & Software Engineering \& Development & \multirow{10}{*}{10} \\
 & Data \& AI & \\
 & Others & \\
 & Security & \\
 & Systems \& DevOps & \\
 & Games \& Entertainment & \\
 & Engineering \& Industry & \\
 & Scientific Research & \\
 & N/A & \\
 & Others & \\
\midrule
\texttt{z2} - Technical domain L2 & A.1 Coding & \multirow{51}{*}{51} \\
 & A.2 Build \& Compilation & \\
 & A.3 Version Control & \\
 & A.4 Web Development & \\
 & A.5 Testing & \\
 & A.6 Debugging & \\
 & A.7 API \& Integration & \\
 & A.8 Document Processing & \\
 & B.1 System Configuration & \\
 & B.2 Network Management & \\
 & B.3 Container \& Virtualization & \\
 & B.4 Logging \& Monitoring & \\
 & B.5 Package Management & \\
 & B.6 File Operations & \\
 & B.7 Automation & \\
 & C.1 Data Processing & \\
 & C.2 Data Analysis & \\
 & C.3 Visualization & \\
 & C.4 ML Training & \\
 & C.5 ML Inference & \\
 & C.6 Computer Vision & \\
 & C.7 NLP & \\
 & C.8 Data Engineering & \\
 & C.9 DL Frameworks & \\
 & D.1 Numerical Computing & \\
 & D.2 Statistical Analysis & \\
 & D.3 Physics Simulation & \\
 & D.4 Bioinformatics & \\
 & D.5 Earth Science & \\
 & D.6 Astronomy & \\
 & D.7 Chemistry & \\
 & D.8 Simulation \& Modeling & \\
 & E.1 Network Security & \\
 & E.2 Application Security & \\
 & E.3 Cryptography & \\
 & E.4 Data Security & \\
 & E.5 Vulnerability Fix & \\
 & F.1 Manufacturing & \\
 & F.2 Energy Systems & \\
 & F.3 Control Systems & \\
 & F.4 Optimization & \\
 & F.5 Planning \& Scheduling & \\
 & F.6 Financial Modeling & \\
 & G.1 Game AI & \\
 & G.2 Game Development & \\
 & G.4 Multimedia & \\
 & H.1 Mathematics & \\
 & H.2 Personal Assistant & \\
 & H.3 General Tools & \\
 & N/A & \\
 & Others & \\
\midrule
\texttt{w} - User role & Software Developer & \multirow{17}{*}{17} \\
 & ML / Algorithm Engineer & \\
 & Operations Product / Content / User / Merchant & \\
 & Data / Business Analyst \& Strategy & \\
 & Ops / SRE & \\
 & Product Manager & \\
 & N/A & \\
 & QA / Test Engineer & \\
 & Others & \\
 & Marketing / Growth & \\
 & UX/UI Designer & \\
 & HR Recruiting & \\
 & Finance / Legal & \\
 & Business / BD / Sales & \\
 & Project Manager & \\
 & Admin Assistant / Secretary & \\
 & PR / Communications & \\
\bottomrule
\end{longtable}
}

\end{document}